\documentclass{article}

\usepackage[utf8]{inputenc}
\usepackage{hyperref}       
\usepackage{graphicx}
\usepackage[numbers]{natbib}
\usepackage{amsmath}
\usepackage{amssymb}
\usepackage{subcaption}
\usepackage{multirow}
\usepackage{booktabs}       
\usepackage[a4paper, total={6.5in, 8.66in}]{geometry}  

\usepackage[linesnumbered,ruled,vlined]{algorithm2e}

\RequirePackage[normalem]{ulem} 
\RequirePackage{color}\definecolor{RED}{rgb}{1,0,0}\definecolor{BLUE}{rgb}{0,0,1} 

\let\oldnl\nl
\newcommand{\nonl}{\renewcommand{\nl}{\let\nl\oldnl}}

\SetCommentSty{mycommfont}

\DeclareMathOperator{\rad}{rad}
\DeclareMathOperator{\bern}{Bern}
\DeclareMathOperator{\bce}{BCE}
\DeclareMathOperator{\sse}{SSE}
\DeclareMathOperator{\sae}{SAE}
\DeclareMathOperator{\dice}{Dice}
\DeclareMathOperator{\pixelsum}{PixelSum}

\newcommand\suep[1]{SUEP(#1\,GeV)}
\newcommand\svj[1]{SVJ(#1\,TeV)}

\title{\huge Triggering Dark Showers with Conditional Dual Auto-Encoders}

\author{
        \href{https://orcid.org/0000-0002-0399-8836}{\includegraphics[scale=0.06]{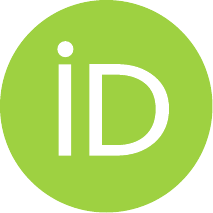}\hspace{1mm}Luca Anzalone}$^{1,3*}$, 
        \href{https://orcid.org/0000-0002-1643-1388}{\includegraphics[scale=0.06]{orcid.pdf}\hspace{1mm}Simranjit Singh Chhibra}$^{1, 2, 5}$,
        \href{https://orcid.org/0000-0001-5270-7540}{\includegraphics[scale=0.06]{orcid.pdf}\hspace{1mm}Benedikt Maier}$^{2,4}$,
        \\
        \href{https://orcid.org/0000-0002-2264-2229}{\includegraphics[scale=0.06]{orcid.pdf}\hspace{1mm}Nadezda Chernyavskaya}$^{2}$,
        \href{https://orcid.org/0000-0003-1939-4268}{\includegraphics[scale=0.06]{orcid.pdf}\hspace{1mm}Maurizio Pierini}$^{2}$
        \\
        \small $^{1}$Department of Physics and Astronomy (DIFA), University of Bologna, Bologna, Italy \\
        \small $^{2}$European Organization for Nuclear Research (CERN), Geneva, Switzerland \\
        \small $^{3}$Istituto Nazionale di Fisica Nucleare (INFN), Sezione di Bologna, Italy\\
        \small $^{4}$Karlsruhe Institute of Technology (KIT), Karlsruhe, Germany\\
        \small $^{5}$Queen Mary University of London (QMUL), London, UK\\
        $^{*}$\small{Corresponding author, email at \texttt{luca.anzalone2@unibo.it}}
        }

\date{}

\hypersetup{
    pdftitle={Triggering Dark Showers with Conditional Dual Auto-Encoders},
    pdfauthor={Luca Anzalone, Simranjit Singh Chhibra, Benedikt Maier, Nadezda Chernyavskaya, Maurizio Pierini},
    pdfkeywords={Anomaly detection, Auto-encoders, Deep learning, Dark showers, High-Energy Physics},
}

\begin{document}

\maketitle

\begin{abstract}
    \noindent We present a family of conditional dual auto-encoders (CoDAEs) for generic and model-independent new physics searches at colliders. New physics signals, which arise from new types of particles and interactions, are considered in our study as anomalies causing deviations in data with respect to expected background events. In this work, we perform a normal-only anomaly detection, which employs only background samples, to search for manifestations of a dark version of strong force applying (variational) auto-encoders on raw detector images, which are large and highly sparse, without leveraging any physics-based pre-processing or strong assumption on the signals. The proposed CoDAE has a dual-encoder design, which is general and can learn an auxiliary yet compact latent space through spatial conditioning, showing a neat improvement over competitive physics-based baselines and related approaches, therefore also reducing the gap with fully supervised models. It is the first time an unsupervised model is shown to exhibit excellent discrimination against multiple dark shower models, illustrating the suitability of this method as an accurate, fast, model-independent algorithm to deploy, e.g., in the real-time event triggering systems of Large Hadron Collider experiments such as ATLAS and CMS. 
\end{abstract}

\section{Introduction}
Model-independent searches are becoming a valid alternative to model-dependent searches at colliders, aiming to discover new physics Beyond the Standard Model (BSM) governed by a vast parameter space~\footnote{A search aims to discard as many background events as possible while preserving the most signal: the background represents what is already well known to exist, i.e., Standard Model (SM) processes or detector effects. The signal, which may or may not exist in Nature, is the object of the search. The scope of analysis is to determine how plausible the existence of a specific signal is.}. Hence, {an} enormous {number of}  model-dependent analyses would {be} required to unravel {such a vast parameter space in its entirety because each analysis need to target a specific signal: in this regard, model-independent searches provide a great, more flexible, alternative.} The conventional cut-based analyses involve physics experts inspecting the distributions of various physical parameters to find discriminating characteristics. Once identified, the best threshold is determined, above which the events are considered signal-like. This part can be automatized by training a Machine Learning (ML) \cite{CMS:2012qbp} or Deep Learning (DL) classifier \cite{Baldi:2014kfa,Anzalone_2022, Kasieczka:2019dbj} separating simulated background and signal events. Subsequently, a rigorous statistical test \cite{Cowan:2010js} determines the significance of the classified signal events: if above a certain threshold, the signal is declared to exist; if too low, the signal can be confidently excluded to exist at all.

Both the cut- and supervised ML-based search techniques are \textit{model-dependent}, i.e., they assume a particular scenario for new physics, thus being signal-specific. For the ML-based approach, the classifier inherently adapts its learned parameters to be sensitive to specific signal features. However, it does not necessarily generalize towards unknown signals. Moreover, a supervised approach requires accurate signal and background simulation and robustness against systematic uncertainties. To mitigate such limitations, we propose a data-driven model-independent search strategy powered by Conditional Dual (Variational) Auto-Encoders (CoDAEs) and normal-only Anomaly Detection (AD) \cite{Heimel:2018mkt}, demonstrating generalization over multiple signals despite not being trained on them.

\noindent In this article, we focus on two important and highly challenging manifestations of Hidden Valley (HV) models \cite{Strassler:2006im}, more specifically, of dark quantum chromodynamics (QCD), namely Soft Unclustered Energy Patterns (SUEPs) \cite{Knapen:2016hky} and Semi-Visible Jets (SVJs) \cite{Cohen:2015toa,Cohen:2017pzm,Kar:2020bws}.
The HV models are a type of BSM physics describing the existence of a new sector of particles and forces with new gauge groups and a mediator to the SM, which the Large Hadron Collider (LHC) \cite{Evans:2008zzb} can produce. 
The HV models have been mainly developed to address the origin of dark matter \cite{Denk:2011sba}, whose experimental signatures often feature non-isolated objects with high-multiplicity and/or low-energy final states, representing a challenging target for existing analyses at the LHC \cite{born2023scouting}. 

Our proposed models can detect both SUEP and SVJ signals in highly sparse raw detector image data, constructed from the trigger system information, within the time budget of the High-Level Trigger (HLT) step \cite{CMS:2016ngn}~\footnote{We considered the Compact Muon Solenoid (CMS) experiment, a general-purpose detector at the LHC \cite{Evans:2008zzb}, as the reference experiment for this study. The CMS trigger system is a two-tiered event selection system. The electronics-based first level (L1) uses information from the calorimeters and muon detectors and reduces the event rate from $40$ MHz to around $100$ kHz within a time interval of $4$ microsec. The second level, known as the High-Level Trigger (HLT), runs a version of the full event reconstruction software optimized for fast processing on a farm of processors. The HLT reduces the event rate to about $1$ kHz within $O(10^2)$ ms, and the selected events are transferred to storage.}, being trained only on the simulated QCD events: the class of data considered as not anomalous. Without making {strong} assumptions about the signals we avoid problem-specific pre-processing, on which discrimination performance can be highly dependent \cite{Kasieczka:2022naq}, and further reduce the dependency on the physics model. Our novel architecture can learn a two-dimensional ({2D}) \textit{auxiliary} latent space through conditioning \cite{dumoulin2018feature-wise}, capturing intrinsic information of the input that can be visualized, interpreted, and, in principle, employed for AD. Our contributions can be summarized as follows:
\begin{itemize}
    \item We frame the new physics search problem as a normal-only anomaly detection task, making {minimal} assumptions on the nature of the signals. {We only assume: 1) to} have access to \textit{normal} (i.e., not anomalous) data samples{, and 2) that the signals can be revealed through tracking information}.
    \item We propose a novel architecture that combines two encoders through spatial conditioning, in order to learn {additional criteria} for discriminating between signal and background.
    \item We perform a comprehensive comparison of anomaly scores, {evaluating} both scores derived from reconstructed images and the latent spaces. 
    \item We ultimately show that our novel auto-encoder can reconstruct the target images with a much higher quality than compared approaches, which can {also} help human experts when visually inspecting anomalies. 
\end{itemize}
\noindent Compared to both weakly-supervised (e.g., \cite{Dillon:2021nxw}) and classification methods (e.g., \cite{Baldi:2014kfa,Anzalone_2022}), which require partial or full knowledge of the signal(s), our approach assumes only the knowledge about the background events. Therefore, potentially enabling generic physics searches for unknown signals. In the following two sections we provide some further physics background relevant to understand our work.

\subsection{The New Physics Search Scenario: Hidden valley models}
The Hidden valley models can produce dark quarks in proton-proton collisions at the LHC, leading to a dark shower and the production of a large number of dark hadrons ($\phi_D$), analogous to QCD jets \cite{Strassler:2006im,Knapen:2016hky}. Depending on the details of the theory, the dark showers can follow large-angle emission and dark hadrons do not form narrow QCD-like jets. The decay of dark hadrons results in dark photons ($Z_D$), which further decay to low-energy SM particles with transverse energy ($E_T$) of $\mathcal{O}(10^2)$\,MeV, whose final experimental signature being high-multiplicity spherically-symmetric soft unclustered energy patterns \cite{Knapen:2016hky}. Through their decay to SM particles via some portal state, like a dark photon, these processes become visible and in principle detectable in $4\pi$-detectors at the LHC such as ATLAS \cite{collaboration2008atlas} and CMS \cite{CMS:2008xjf}. We focus on a well-motivated scenario where SUEP is produced in exotic Higgs ($H$) boson decays via gluon-gluon fusion and all dark hadrons decay promptly and exclusively to pions and leptons, an experimental nightmare scenario because of an overwhelming multi-jet QCD background.

Another manifestation of hidden valleys can be semi-visible jets \cite{Kar:2020bws,Cohen:2015toa}, a phenomenon in which energetic particles are emitted in a spray of stable invisible dark matter along with unstable states that decay back to SM. These showers are partially detectable, with the visible components looking like QCD showers \cite{Cohen:2015toa}. This partial visibility makes it challenging to identify and study these particles thoroughly, having a low acceptance with current methods. 

\subsection{The CMS Detector and simulated samples}
\begin{figure}[h]
    \centering
    \includegraphics[width=0.85\textwidth]{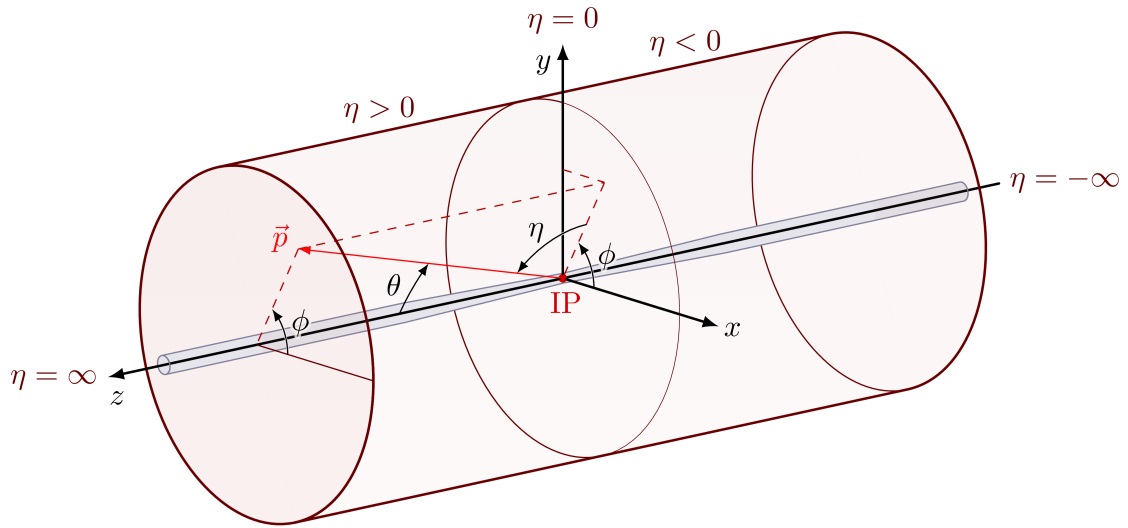}
    \caption{The CMS coordinate system, which explains the particle's motion within the cylindrical detector. Figure adapted from \url{https://tikz.net/axis3d_cms/}.}
    \label{fig:cms_coord}
\end{figure}

The CMS experiment \cite{CMS:2008xjf} is designed to explore the physics of proton-proton collisions through a system of different sub-detectors, each designed to measure different aspects of the particles produced in a collision. Given its \textit{cylindrical} design, as we can see in figure \ref{fig:cms_coord}, it is often convenient to adopt a polar coordinate system $(\theta, \phi)$ where: $0\le \theta\le \pi$ is the polar angle, and $0\le\phi\le 2\pi$ is the azimuthal angle. From these coordinates, it is possible to explain the particle's kinematic as $(p_T,y,\phi, m)$: where $m$ is the invariant mass, $p_T$ the transverse momentum, and $y$ the rapidity. A quantity related to the rapidity is the \textit{pseudo-rapidity} $\eta$, which is a measure of the angle of the particle's motion relative to the beam line. The images employed in our study are represented in the $\eta$-$\phi$ plane, therefore considering the pseudo-rapidity and azimuth.

\noindent The CMS detector consists of several layers that are used to measure various properties of particles produced in high-energy collisions. The ones \cite{CMS:2017yfk} relevant to our work are the:
\begin{itemize}
    \item \textbf{Inner tracking system}, which measures the momentum of particles by their curvature radius through the magnetic field. The tracker can monitor the paths of charged particles. This sub-detector covers a pseudo-rapidity region of up to $|\eta| < 2.5$, being made of $66$M silicon pixel detectors ($100\times 150 \mu m^2$ in size) for accurate measurement of the particle's trajectory.
    \item \textbf{Calorimeters}, consisting of an Electromagnetic calorimeter (ECAL), and a Hadron calorimeter (HCAL). The calorimeters can measure the direction and energy of both charged and neutral particles. The two sub-detectors have different granularity: for the ECAL, the granularity of $0.0174\times 0.0174\rad^2$ results in $286\eta\times 360\phi$ bins for the size of the images, whereas the HCAL is 25 times less granular, i.e., $0.087\times 0.087\rad^2$. Therefore, each HCAL image is upsampled by a factor of 25 in the preprocessing step, giving $1/25$th of the energy to each pixel.
\end{itemize}

\noindent We employ the Delphes v.3.4.3pre1 fast detector simulation~\cite{delphes} with the CMS Run-2 detector model to obtain the tracker, ECAL, and HCAL images. Samples of SM multijet events as well as for SVJ and SUEP signal processes have been generated with the Pythia v8.244 event generator~\cite{Sjostrand:2014zea}.

\section{Related Work}
\noindent In this section we review the relevant literature about anomaly detection in high-energy physics (HEP). AD \cite{DBLP:journals/corr/abs-1901-03407} is the task of determining which samples violate some notion of normal behavior: once identified, {these} samples will be referred to as \textit{outliers} or \textit{anomalies}. We assume a normal-only setting{, in which the training is performed only on background data representing the already known (i.e., not anomalous) behavior}, being a good approximation to what occurs in practice, i.e., having the background contaminated with a little fraction of unknown signals. (Variational) Auto-Encoders (V/AEs) \cite{DBLP:journals/corr/KingmaW13, hinton2006reducing, DBLP:conf/icann/MasciMCS11} are a popular mean to perform AD: the model is trained to minimize the reconstruction error of the normal samples, which is then used to score the novel data. Anomalies are found by thresholding such error. A general challenge is about designing anomaly scores that best separate the normal data from the anomalies \cite{Fraser:2021lxm}; to this end, V/AEs allow conceiving two main classes of anomaly scores, as described in the next two sections.

\subsection{Reconstruction-based Anomaly Detection}
\noindent Reconstruction-based anomaly scores are obtained by comparing the reconstructions, $\hat{x}$, with the inputs, $x$, of the V/AE. Different scores can be determined according to the distance or similarity function used to compare images. Heimel et al. \cite{Heimel:2018mkt} introduce the benchmark dataset of QCD vs top jets. Their approach heavily relies on a specific particle-based processing of the raw collisions, which greatly simplifies the problem. Their LoLa AE, which is based on jet-level kinematics features, is able to beat an image-based AE by a large margin even with a smaller latent space, although at the cost of introducing an even larger dependency on the jet mass. Finke et al. \cite{Finke:2021sdf} discuss the limitations of using AEs on the same kind of data. The authors propose the kernel-MSE loss function, which is less sensitive than MSE, encouraging the network to learn dim pixels even in presence of sparsity. Recently, Dillon et al. \cite{dillon2023normalized} propose to use a normalized auto-encoder (NAE) \cite{DBLP:conf/icml/YoonNP21} to identify anomalous jets symmetrically. The NAE maximize the likelihood of the data through the minimization of an energy function. Under this probabilistic formulation, the NAE is forced to inhibit the reconstruction of an outlier, since it has to maximize the likelihood of the normal data, guaranteeing a low reconstruction error only for them. Although, NAEs are well-suited for anomaly detection, avoiding their training instabilities is still a practical challenge.

\subsection{Latent-based Anomaly Detection}
\noindent Latent-based anomaly scores are defined from the latent space captured by the encoder network: directly {using} the learned {latent representation} to flag anomalies can be difficult due {its} high-dimensionality, therefore combining the information carried by each {latent} component may require explicit supervision \cite{metodiev2017classification,DBLP:conf/iclr/HendrycksMD19}. Dillon et al. \cite{Dillon:2021nxw} proposed to use a Dirichlet VAE \cite{DBLP:journals/pr/JooLPM20} to learn a bi-modal, one-dimensional latent space that naturally encodes the two classes{: signal and background}. The authors show that the Dirichlet prior on the latent space naturally leads to mode separation, something that was not observed for both the regular VAE \cite{DBLP:journals/corr/KingmaW13} and the Gaussian-mixture VAE \cite{DBLP:journals/corr/DilokthanakulMG16}, without enforcing any additional loss term. The proposed Dirichlet VAE reaches high class separation performance although weak-supervision is still required. Bortolato et al. \cite{Bortolato:2021zic} propose to use the Kullback-Leibler divergence (KLD) between the learned and prior Gaussian distributions as an anomaly score to detect anomalous jets. Dillon et al. \cite{Dillon:2022tmm} compared the effectiveness of using low-dimensional latent space representations instead of the event space features to perform model-agnostic anomaly detection. They trained a transformer encoder \cite{DBLP:conf/nips/VaswaniSPUJGKP17} to optimize the JetCLR's contrastive objective \cite{Dillon:2021gag}, where symmetry augmentations were employed to define positive and negative pairs for the contrastive learning. Through a binary classification test, the authors {discovered} that a sufficiently large latent space (e.g., of size $512$) is required to encode the physical symmetries of jets. Finally, the CWoLa \cite{metodiev2017classification} method was used to perform model-agnostic anomaly detection, showing that still a significant fraction of signal events is required to achieve meaningful class separation. Govorkova et al. \cite{Govorkova:2021utb} demonstrate a real-world deployment of a VAE on FPGA hardware for real-time AD at the LHC \cite{Evans:2008zzb}. The authors compared the performance of both reconstruction- and KL-based anomaly scores, for both AE and VAE models. They concluded that with a minor loss in performance, the scores based on the KL divergence allowed them to only deploy the VAE's encoder on the FPGA, thus saving both hardware resources and latency costs. Recently, Cheng et al. \cite{Cheng:2020dal} enhanced a VAE with a technique known as Outlier Exposure (OE) \cite{DBLP:conf/iclr/HendrycksMD19}, which makes use of an auxiliary set of out-of-distribution (OOD) data to improve the sensitivity to anomalies. An auxiliary loss term is computed from OOD predictions, which ensured a good compromise between high separation of anomalies and jet mass decorrelation. Although the promising results, it is not yet clear if data from the same physics domain is enough to be considered as OOD.

\subsection{Related High-Energy Physics Analyses}
\noindent The analyses conducted in \cite{Barron:2021btf} and \cite{Canelli:2021aps} are related to ours, since it is assumed a similar signal setting. In particular, Barron et al. \cite{Barron:2021btf} target the same SUEPs scenario in which the signal decays to exotic Higgs, and all the dark hadrons to SM hadrons. The authors identify three observables: charged particle multiplicity, event isotropy, and interparticle distance. These are used to build the input features for their unsupervised fully-connected auto-encoder. Canelli et al. \cite{Canelli:2021aps}, instead, study the SVJs signature by training a fully-connected auto-encoder on jet-level and jet substructure variables, minimizing the mean absolute error. Compared to these two studies, we neither rely on high-level nor engineered {particle-based} features but, instead, learn from {raw} detector images. Moreover, our models are evaluated against both signals, demonstrating anomaly scores that can identify both.

\section{Simulated Dataset of particle collisions}
\noindent The dataset employed for our study contains simulated images of size $360\times 286\times 3$, for a total of about $615$k samples, divided in: $442$k QCD, $67$k SUEPs, and $106$k SVJs.
The image channels represent two-dimensional $E_T$ (energy) deposits in the $\eta$-$\phi$ plane,
which are measured by the Inner tracker (Trk), ECAL, and HCAL sub-detectors \cite{CMS:2017yfk} of CMS \cite{CMS:2008xjf}, respectively. 
Moreover, each image is annotated with a:
\begin{itemize}
    \item \textbf{Class label.} There are three of them in total: the label $0$ indicates the QCD background, the label $1$ is associated to SUEP signal samples, and the label $2$ refers to the second SVJ signal.
    \item \textbf{Mass label.} Signal samples only are identified by the mediator masses $m_H$~\cite{Knapen:2016hky} and $m_{Z'}$~\cite{Canelli:2021aps} at which these were generated. In particular,  SUEPs were generated at $m_H = \{125, 200, 300, 400, 700, 1000\}$\,GeV and SVJs at $m_{Z'}=\{2.1, 3.1, 4.1\}$\,TeV. For the rest of the paper, we refer to a particular signal sample by its mediator mass, such as \suep{$m_H$} and \svj{$m_{Z'}$}.  
    \item \textbf{Number of tracks.} This is a model-independent quantity that best approximates the number of decay products, obtained by the particle-flow reconstruction algorithm \cite{CMS:2017yfk}. We refer to this variable as \verb|nTracks|. It should be noticed that computing this quantity is expensive, being not feasible for real-time inference at the HLT.
\end{itemize}

\noindent Since the images are very sparse, having about $99.4\%$ of zero pixels, and also moderately large, we employ a simple pre-processing {(as described by algorithm \ref{algo:preproc})} that down-scales the images, thus reducing sparsity while also preserving their total energy. The down-scaling is performed by convolving a $5\times 5$ kernel with all ones on the input images along the channel dimension (i.e., in a depth-wise fashion), in a non-overlapping manner with a stride equal to the kernel size, yielding a $25\times$ reduction in spatial resolution while preserving the sum of the energy deposits: the sparsity is also reduced {to} $96\%$; {zero-padding is also applied to let the output size be divisible by the kernel size.} The last step of the pre-processing is to discard the HCAL and ECAL channels, considering only the tracker one, resulting in images of size $72\times 58\times 1$: smaller images are faster to process by the network and {require less storage}, allowing to save up parameters, memory, computation and time.

\begin{algorithm}[!ht]
    \DontPrintSemicolon
  
    \nonl \textbf{Input:} a batch of images $I\in\mathbb{R}^{B\times H\times W\times C}$, kernel size $K$\\
    \nonl \textbf{Output:} pre-processed images $I_M^{trk}\in\mathbb{R}^{B\times \lceil H/K\rceil \times \lceil W/K\rceil\times 1}$\\
    
    \tcc{Depthwise convolution to down-sample each channel by a factor of $K$}
    kernel = \texttt{tf.ones}(($K,K,C,1$))\\
    $I'$ = \texttt{tf.nn.depthwise\_conv2d}($I$, filter = kernel, strides = ($1,K,K,1$), padding = "SAME")
    
    \tcc{Consider only the tracker channel, discarding the other two}
    $I^{trk}$ = $I'$[$\ldots,\ 0,$ \texttt{tf.newaxis}]
    
    \tcc{Compute the \textit{mask} image}
    $I_M^{trk}$ = \texttt{tf.cast}($I^{trk} > 0$, dtype = float)
    
    \BlankLine
    \KwRet{$I_M^{trk}$}
    
    \caption{Image Pre-processing}
    \label{algo:preproc}
\end{algorithm}

\section{Method}
\noindent In this section we detail our CoDAE {architecture and training procedure, the physics-inspired image augmentations applied to the pre-processed input images,} and also define a variety of reconstruction- and latent-based anomaly scores.

\subsection{Image feature-engineering}
\begin{figure}
    \centering
    \includegraphics[width=1.0\linewidth]{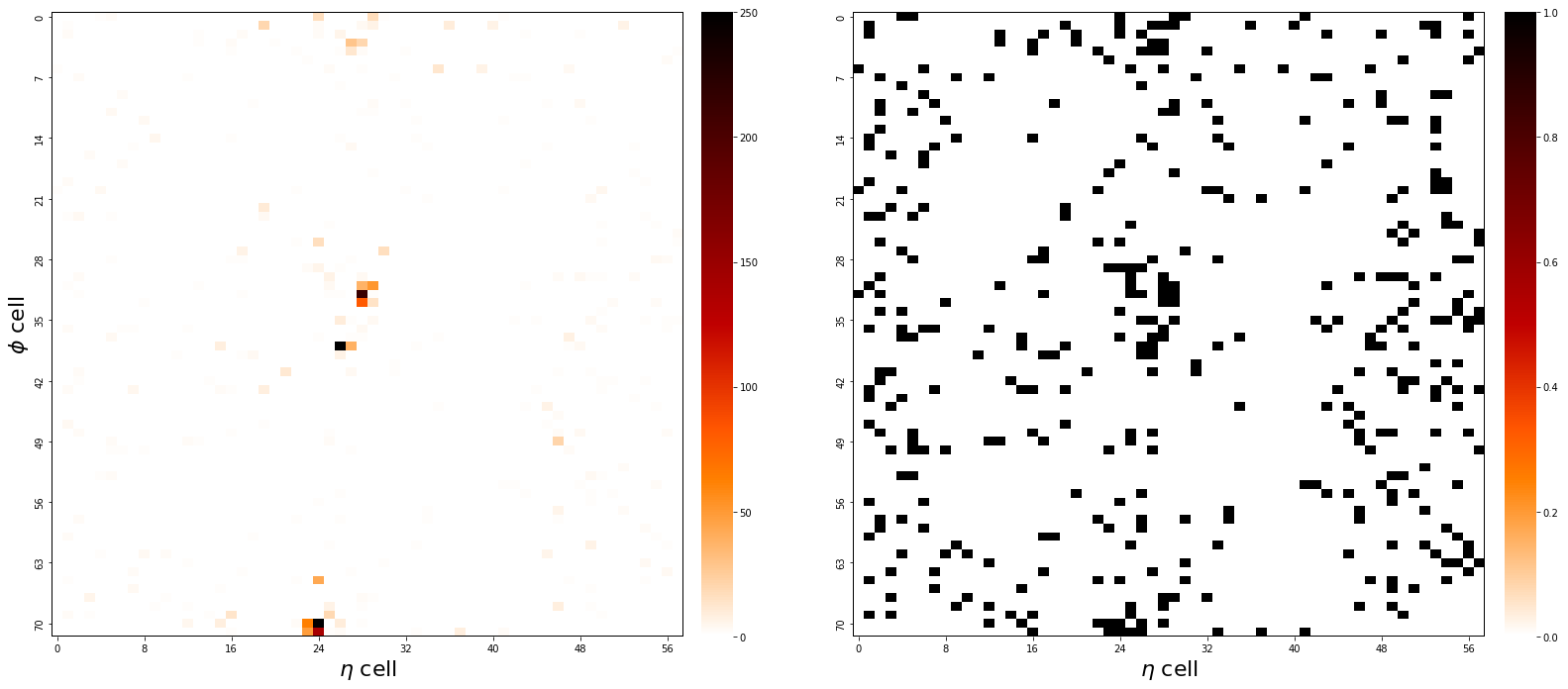}
    \caption{Energy, $I$, (left) and mask, $I_m$, (right) images of a single sample. Tracks information turned to be more discriminative than energy deposits, so we train on $I_m$ instead of $I$.}
    \label{subfig:mask_img}
\end{figure}

\noindent Both the energy deposits and the \verb|nTracks| variable can be seen as physics-motivated discriminators. Moreover, the number of tracks is much more sensitive to the searched signals than the energy, {as stated in \cite{Knapen:2016hky} and confirmed by our prior experiments}, representing a better input for our models. Therefore, we devised a simple way to approximate the \verb|nTracks| information by "feature-engineering" the energy images, $I$, where each pixel depicts an $E_T$ deposit, without running track reconstruction algorithms. 
The resulting \textit{mask image}, $I_m$, is obtained by determining whether a pixel depicts a non-zero energy value: $I_m = \mathbf{1}[I > 0]$, where $\mathbf{1}[\cdot]$ is an indicator function applied to each pixel of $I$.
Each pixel in $I_m$ represents whether or not a single track has occurred, so its value can be at most one{: a comparison of both kinds of images is shown in figure \ref{subfig:mask_img}.}

A mask image, if summed, denotes the number of non-zero deposits associated with sensors in the detector that measured some energy. {This} quantity is similar to the \verb|nTracks| variable, but not equivalent since, depending on the granularity used to yield the images, two or more tracks can fall in the same bin (pixel) thus being not distinguished when counting non-zero pixels. In particular, we consider the mask image computed from the energy deposits of the tracker channel only, as the calorimeter information turned to be not enough informative: this fact was validated by our prior experiments, in which one possible explanation provided by \cite{Knapen:2016hky} is that at the calorimeter level the SUEP resembles the pile-up since lacking hard and isolated objects, therefore, being more noisy than informative. Moreover, pixels in $I_m$ have the additional benefit of being either zero or one, avoiding the need of normalizing $E_T$ deposits which are often large in range and skewed towards small values.

\subsection{Image Augmentations}
\label{subsec:data_aug}
\noindent Since our data have physical properties like total energy, and the $E_T$ deposits are arranged according to the design of the detector (other than being the result of a physics phenomenon), we cannot simply apply the usual off-the-shelf image augmentations like random crop, cutout, rotation, and jittering \cite{DBLP:journals/jbd/ShortenK19} that would reorganize the image' pixels without following the underlying physics and also without preserving both the individual and overall value of {energy} deposits. For {this} reason{,} we design novel data augmentations that preserve the physical meaning of the images {by} working on the $\eta$-$\phi$ plane{, hence,} also respecting the {geometry of the detector}.

In particular, one kind of image augmentation involves a flipping in $\eta$, while the other is a rotation in $\phi${: considering the figure \ref{fig:cms_coord} as a visual reference, the former can be interpreted as considering particles moving in the opposite direction with respect to the beam line (i.e., along the detector's $z$-axis), whereas the latter as rotating clockwise or anticlockwise the whole collision event on the detector's $x$-axis.} The $\eta$-flip augmentation {is} simply implemented by mirroring the $x$-axis from left to right, where the $\phi$-rotation is a little more complex. Rotation in $\phi$ (i.e., along the {image's} $y$-axis\footnote{We refer to Cartesian $x$ and $y$ axes in the context of images, rather than $x$ and $y$ (i.e., rapidity) as detector coordinates.}) can occur both upward {(i.e., anticlockwise)} and downward {(clockwise)}, in which a portion of the image moves up (or down) and the part in excess (the one that would fall off vertically from the image boundaries) is then attached to the bottom (or top). From a practical perspective, the $\phi$-rotation is done in chunks of $\Delta$ rows, {in which} the chunk size {is uniformly sampled for each image that should be rotated from the set}, {$\Delta\in\{8,16,\ldots,56\}$}, {whose values} are only multiples of eight: a hyperparameter value found to work well experimentally.

\begin{figure}[h!]
    \centering
    
    \begin{subfigure}{0.32\textwidth}
        \includegraphics[width=\textwidth]{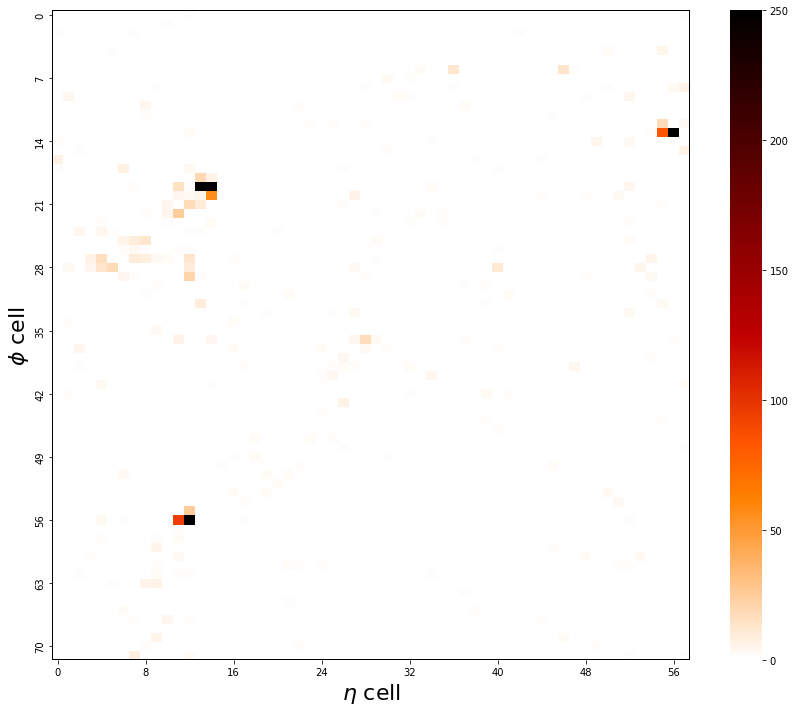}
        \caption{No augmentation}
    \end{subfigure}
    \hfill
    \begin{subfigure}{0.32\textwidth}
        \includegraphics[width=\textwidth]{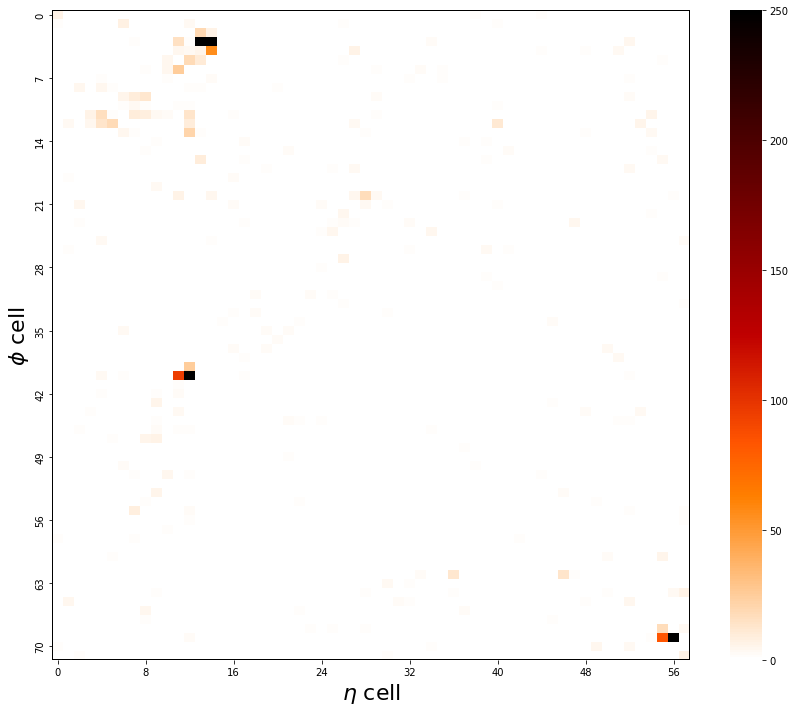}
        \caption{Upward rotation in $\phi$ ($\Delta = 16$)}
    \end{subfigure}
    \hfill
    \begin{subfigure}{0.32\textwidth}
        \includegraphics[width=\textwidth]{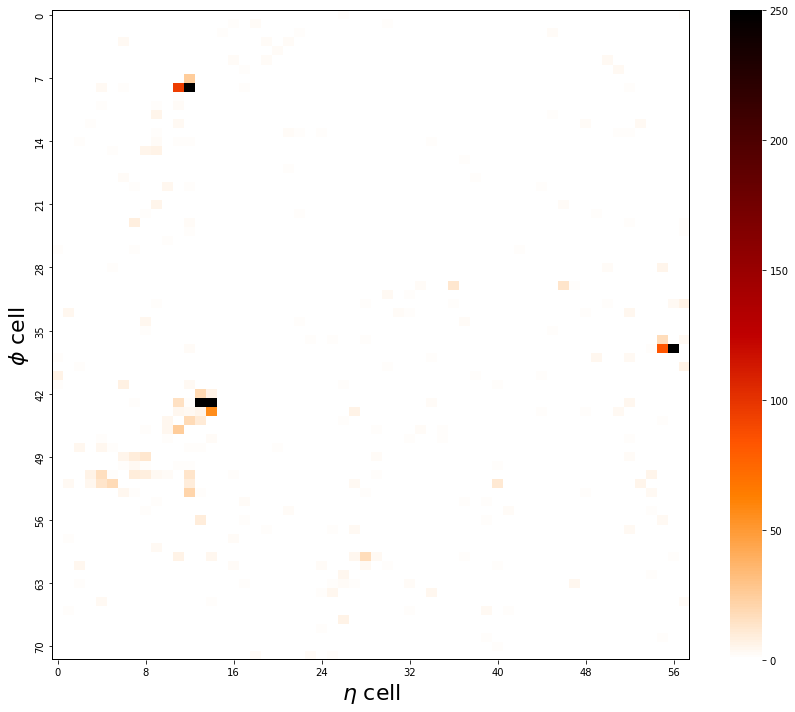}
        \caption{Downward rotation in $\phi$ ($\Delta = 24$)}
    \end{subfigure}

    \begin{subfigure}{0.32\textwidth}
        \includegraphics[width=\textwidth]{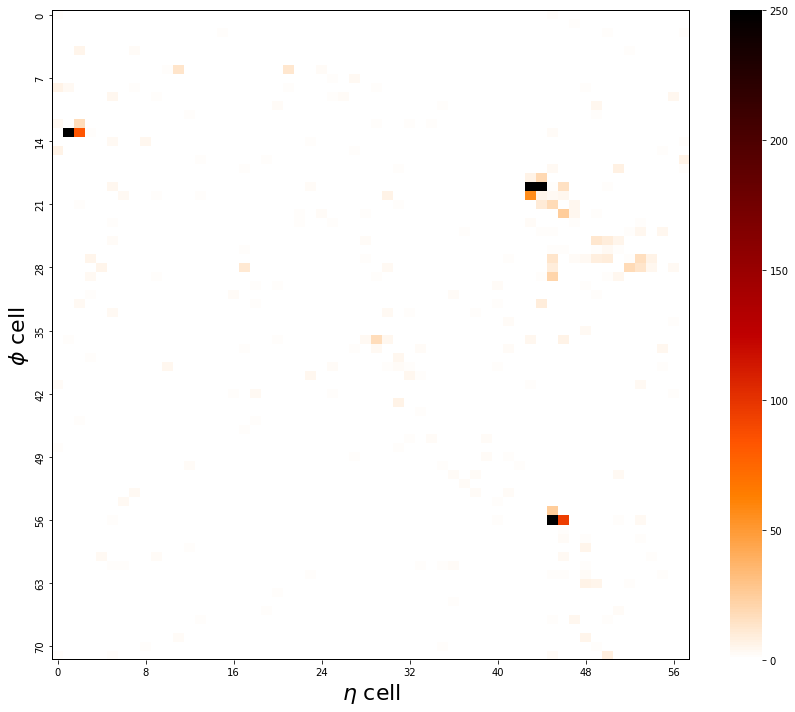}
        \caption{Flip in $\eta$}
    \end{subfigure}
    \hfill
    \begin{subfigure}{0.32\textwidth}
        \includegraphics[width=\textwidth]{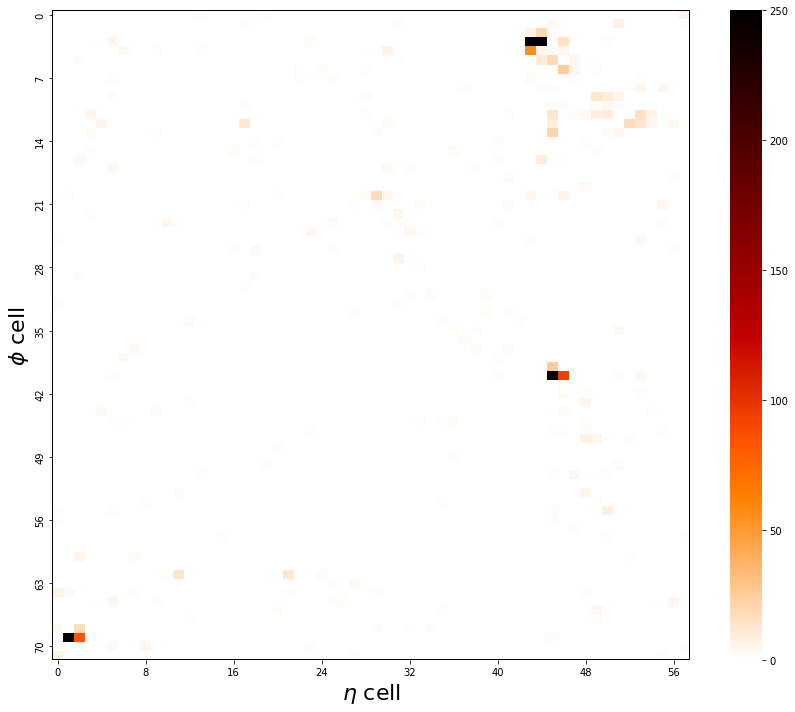}
        \caption{Flip and upward rotation}
    \end{subfigure}
    \hfill
    \begin{subfigure}{0.32\textwidth}
        \includegraphics[width=\textwidth]{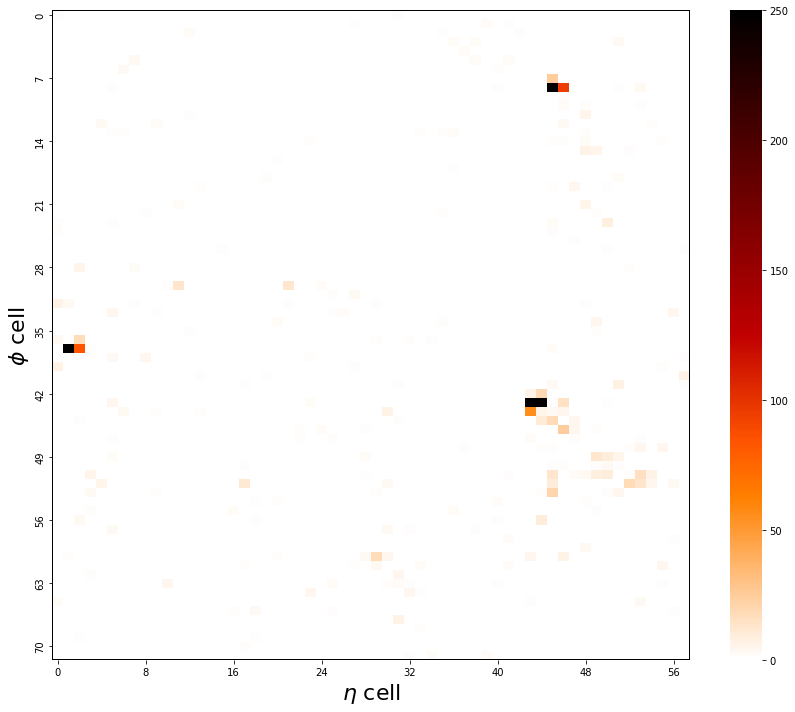}
        \caption{Flip and downward rotation}
    \end{subfigure}
    
    \caption{Example of five kinds of data augmentations, demonstrated on a random QCD sample {considering energy deposits, whose value scale is depicted by the colored bar}.}
\label{fig:data_aug}
\end{figure}

\noindent By combining these two kinds of image augmentations, flipping and rotation, it is possible to yield a total of five combinations of augmentations: 1) upward rotation, 2) downward rotation, 3) flipping, 4) flipping and upward rotation, and lastly 5) flipping with downward rotation. {The image augmentations presented in figure \ref{fig:data_aug} can be applied to raw images of particle collisions, regardless the specific kind of signal and background processes. Moreover, these are also designed to encourage the model to be invariant with respect the detector geometry: learning the properties of the detector's coordinate space is useful not only for anomaly detection but also for classification and regression problems.} 
 
\subsection{The Dual Encoders}
\label{subsed:dual_enc}
\noindent Capturing a latent space that is both discriminative for anomaly detection, and high-capacity for accurate pixel-level reconstructions can be challenging due to a trade-off between the size of the {latent space} and the reconstruction quality. Large latent spaces yield accurate reconstructions but cannot be used directly as AD scores unless summarized in some way. Conversely, small {latent spaces} can encode discriminative features but at the cost of poor reconstructions due to the low-dimensionality that does not retain pixel-level details{: this fact can stop the training prematurely, resulting in sub-optimal anomaly scores that may not even discriminate the anomalies}. {Since the model is trained to maximize the reconstruction fidelity, ensuring a good convergence of the loss can indirectly improve the anomaly scores too since these are, even if not optimized directly, defined from the reconstructions: reaching a better optimum entails a lower reconstruction loss on the normal examples, which, in the context of AEs, means obtaining a better (i.e., more structured, rich) latent space and a more accurate decoder.}

As we want to capture both the detail and discriminative features, we define two encoders, $f_R$ and $f_m$, trying to disentangle these two notions without any additional supervision. The encoder $f_R$ is a {residual network \cite{DBLP:conf/cvpr/HeZRS16}} {that embeds the input images in a large convolutional latent space}, $Z = f_R(I_m)$, {of size $|Z|=5\times 4\times 64$: given its large capacity, the latent components are expected to retain enough information to let the decoder yield high-quality reconstructions.}
The mask encoder $f_m$, instead, is a shallower convolutional network aimed at learning a compact and discriminative auxiliary latent space{,} $Z_m = f_m(I_m)$ {where} $|Z_m| = 2${, such that its components can be directly used as anomaly scores}. Both encoders receive the same mask image, $I_m$, as input. Furthermore, the two networks have different architectures to induce a bias during training, established with prior experiments: $f_R$ is high capacity and its skip connections can propagate information deeply in the layers' hierarchy helping to retain {pixel-level} details, whereas the max pooling layers in $f_m$ are meant to consider only the most important activations to enhance the discrimination power of $Z_m$. {For clarity, we also refer to $Z$ as the \textit{convolutional latents}, and $Z_m$ as the \textit{auxiliary or compact latents}.} 
 
\subsection{The Conditional Decoder}
\begin{figure*}
    \centering
    \includegraphics[width=0.99\textwidth]{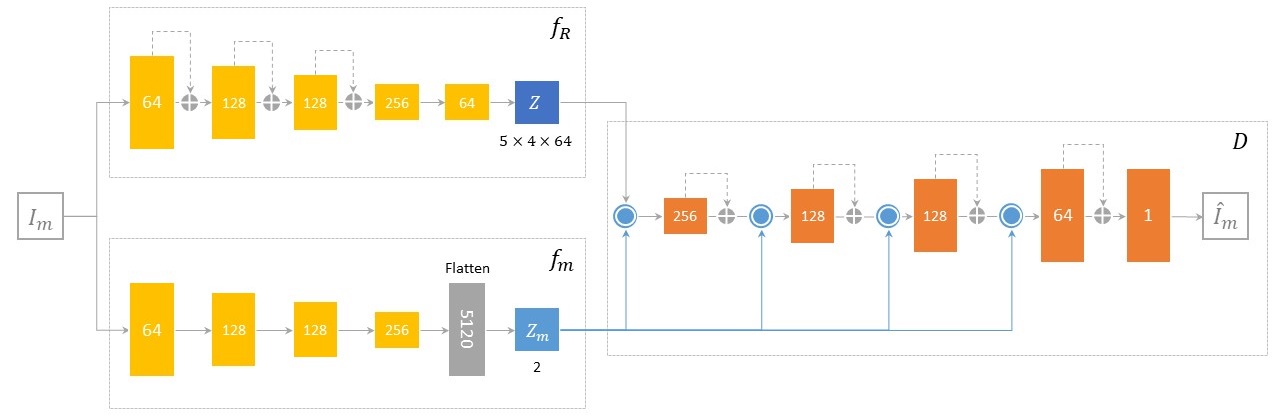}
    \caption{The architecture of both CoDAE and CoDVAE: convolutional and upsampling blocks in yellow and orange, respectively, gray dashed arrows and a $\oplus$ symbol depict a skip (or residual) connection \cite{DBLP:conf/cvpr/HeZRS16}, lastly blue solid arrows and the $\circledcirc$ symbol denote spatial conditioning of $Z_m$ on $Z$. Each block uses $3\times 3$ convolutions followed by instance normalization \cite{DBLP:journals/corr/UlyanovVL16} and Leaky-ReLU \cite{maas2013rectifier}. Down-sampling in the encoder is performed by strided convolutions instead of pooling.}
    \label{fig:codae_arch}
\end{figure*}

\noindent The conditional decoder $D$ is a residual {network} \cite{DBLP:conf/cvpr/HeZRS16} whose main input is the convolutional {latent space}, $Z$ (i.e., the output of the residual encoder, $f_R$), from which it tries to reconstruct the input mask images, $I_m$. The {latent space}, $Z$, {is sufficiently large to} provide enough information {to the decoder} to enable high-quality reconstructions; but the question is about how to enable the mask encoder, $f_m$, to learn a compact latent space, $Z_m$. The answer is provided by \textit{conditioning} \cite{dumoulin2018feature-wise, DBLP:journals/corr/MirzaO14}, which establishes a dependency between the decoder, $D$, and the auxiliary latent space, $Z_m$, allowing the gradients of the loss to flow through $f_m$ without any direct supervision. We call the whole auto-encoder architecture a \textit{Conditional Dual Auto-Encoder} (or CoDAE): described in figure \ref{fig:codae_arch}.

During training, the conditioning mechanism propagates the reconstruction error also to the mask encoder, {without any additional loss term or extra supervision,} providing feedback to learn $Z_m$ such as to maximize the reconstruction quality. Turns out that $Z_m$ alone is not enough for pixel-accurate reconstructions\footnote{Even with a high-capacity residual encoder, reconstructing from only the two components of $Z_m$ results in reconstructions that look like just \textit{average images}, thus without pixel-level details.}, and so we also need to learn the high-capacity latent space, $Z$. Intuitively, we write $\hat{I}_m = D(Z\mid Z_m)$ to highlight that the compact {latents}, $Z_m$, must influence the reconstructions, $\hat{I}_m$, in order to represent meaningful and not just random encoded features. For such reason, the conditioning should be strong enough to prevent $D$ from completely relying only on the {convolutional latents}: in its base form, $Z$ is modulated by {conditional scaling} \cite{dumoulin2018feature-wise} {(i.e., a conditioning mechanism that establish a dependency through multiplications)}, which occurs at multiple levels of the decoder hierarchy.

In particular, our form of conditioning combines \textit{spatial broadcast} \cite{DBLP:journals/corr/watters-abs-1901-07017} with a feature-wise transformation \cite{dumoulin2018feature-wise}: element-wise multiplication or scaling. We will refer to {this} operation as \textit{spatial conditioning}: described by equation \ref{eq:spatial_cond} and algorithm \ref{algo:spatial_cond}. The spatial broadcast ($\text{SB}$) operation provides an inductive bias to the convolutional encoder, $f_m$, for learning \textit{disentangled} latent factors in $Z_m$ (which should encourage to capture independent features over the latent components) while, at the same time, modulating $Z$ and the subsequent hidden feature maps. Spatial conditioning is performed at multiple spatial {resolutions} of the decoder's hierarchy of layers. Initially, at stage $i=0$, the conditioning is performed on {the convolutional latents} (i.e., $h_0=Z$) which {are} then fed to the first layers of the decoder. Subsequently ($i>0$) the hidden feature maps (output of the previous residual block in stage $i-1$), $h_i$, are conditioned on the same $Z_m$. In this way, the {auxiliary} latent space effectively modulates the decoder at different spatial resolutions. The operation performed on a generic tensor $h_i$, of size $H_i\times W_i\times C_i$, can be written as:
\begin{equation}
    r_i = \text{Conv}\big(\text{SB}(Z_m) \big) \odot h_i,
    \label{eq:spatial_cond}
\end{equation}
where $\text{SB}$ (see algorithm 1 in \cite{DBLP:journals/corr/watters-abs-1901-07017}) replicates and expands $Z_m$ to match the shape of $H_i\times W_i$, the subsequent convolution ($\text{Conv}$) linearly expands the channels of the intermediate result to $C_i$, to finally perform the Hadamard product (denoted by $\odot$) with $h_i$, yielding the conditioned representation $r_i$ at stage $i$. 

\begin{algorithm}[!ht]
    \DontPrintSemicolon
  
    \nonl \textbf{Input:} latents $Z_m\in\mathbb{R}^2$, tensor $h_i\in\mathbb{R}^{H_i\times W_i\times C_i}$, kernel size $K$\\
    \nonl \textbf{Output:} spatially conditioned representation $r_i\in\mathbb{R}^{H_i\times W_i\times C_i}$\\
    
    \tcc{See algorithm 1 in \cite{DBLP:journals/corr/watters-abs-1901-07017}}
    $z$ = \texttt{SpatialBroadcast}($Z_m$, $W_i, H_i$)
    
    \tcc{Expand channels of $z$ to match $h_i$, through a linear convolution}
    $z$ = \texttt{Conv2D}(filters = $C_i$, kernel\_size = $K$, padding = 'same')($z$)
    
    \tcc{Multiplicative conditioning: Hadamard product $z\odot h_i$}
    $r_i$ = \texttt{tf.multiply}($z$, $h_i$)
    \BlankLine
    \KwRet{$r_i$}
    
    \caption{Spatial Multiplicative Conditioning}
    \label{algo:spatial_cond}
\end{algorithm}

\noindent In general, our spatial conditioning operation is not limited to only multiplicative interactions. Other simple conditioning mechanisms applied on feature maps are possible, like addition ({also called} biasing), concatenation, or even an affine-like operation that combines both multiplication (also known as scaling) and biasing. {In principle, it is also possible to exploit the domain knowledge of the problem and data to devise an application-specific conditioning mechanisms that utilizes such knowledge: for example, we may think of conditioning on a relevant physics observable.}

\subsection{Categorical CoDVAE}
\noindent The described CoDAE architecture can be easily extended to Variational Auto-Encoders \cite{DBLP:journals/corr/KingmaW13} (VAE). In particular, we explored the Categorical VAE \cite{DBLP:conf/iclr/MaddisonMT17,DBLP:conf/iclr/JangGP17} in which the {convolutional} latents, $Z$, are sampled from a Gumbel-Softmax {(or Concrete)} distribution: $Z\sim Cat(\alpha_R, \tau_R)$, where $\alpha_R,\tau_R = f_R(I_m)$ are the learned logits and temperature. The motivation is that the Categorical VAE can learn discrete features, like counts, whereas the regular (i.e., Gaussian) VAE captures continuous quantities which are not completely related to the \verb|nTracks| variable. We call this model the \textit{Categorical Conditional Dual Variational Auto-Encoder}, or CoDVAE in short. Likewise, in previous approaches \cite{DBLP:conf/nips/HavrylovT17,DBLP:journals/spic/YanSLZ18}, we let the residual encoder also output a temperature $\tau_R$, which can be difficult to tune properly, defining the degree of relaxation of the Categorical distribution {(approximated by the Gumbel-Softmax)} parameterized by $\alpha_R$. Basically, we add a second point-wise convolution with a softplus activation, to ensure positive values, on the last residual block of figure \ref{fig:codae_arch} (diagram section about $f_R$). Then, a base temperature, $\tau_0$, which is a hyperparameter set to $1$, is added to $\tau_R$ to avoid gradient instabilities {due to small numbers} as $\tau_R\to 0$ recovers the true Categorical distribution{: conversely, if $\tau_R\to\infty$ the distribution approaches a Uniform one}. Since we learn the temperature, we define a more complex prior $p(Z)$ as a uniform mixture of $N$ Gumbel-Softmax with different temperatures sampled uniformly in $[0.1, 1]$, to provide more flexibility to the latent space:
\begin{equation}
    p(Z) = \frac{1}{N}\sum_{i}^N Cat(\log 1/C, \tau_i),\quad \tau_i\sim U(0.1, 1).
\end{equation}
We then approximate the KL divergence ($D_{KL}$) between the prior and the learned distribution, $q=Cat(\alpha_R,\tau_R)$, with a Monte-Carlo estimate ($M=10$), as follows:
\begin{equation}
    D_{KL}(q\ \|\ p) \approx \frac{1}{M}\sum_{z\sim q(x)}^M\log q(z\mid x) - \log p(z).
\end{equation}
To enable sharp reconstructions of mask images, we opt for learning a probabilistic decoder in which each output pixel is independently governed by a Bernoulli distribution, which is able to represent pixel values that are either zero or one, as in $I_m$. A single Bernoulli distribution can be denoted as $\bern(p_\theta)$ where $p_\theta$ is learned and specific for a single pixel. For sharp predictions we take the \textit{mode} of such distribution since it is $1$ if $p_\theta > \frac{1}{2}$ and $0$ otherwise: its mean, instead, is suitable to represent smooth values in $[0, 1]$ like normalized energy deposits.


\subsection{Anomaly Scores}
\label{subsec:anomaly_scores}
\noindent Auto-encoders provide the opportunity to define a variety of scores to perform anomaly detection. Our CoDAE architecture learns an auxiliary low-dimensional latent space, allowing each component to be an anomaly score: in our case $|Z_m|=2$, so we have two discriminators\footnote{In principle, the latent space $Z_m$ could have an arbitrary number of dimensions but there is a trade-off: learning either too numerous or too few (e.g., one) latent components may result in individual scores with a weak discriminatory power.}. In addition to this, VAEs provide a natural way to discriminate on the latent space through the KLD \cite{Bortolato:2021zic,Govorkova:2021utb} between the learned posterior and the prior distribution, i.e., $D_{KL}\big(q(z\mid x)\|\ p(z) \big)$. Yet another option is provided by the fact that the KLD is not \textit{symmetrical}: $D_{KL}(q\ \|\ p)$ and $D_{KL}(p\ \|\ q)$ have two different meanings. The former promotes mode-seeking behavior (called the \textit{reverse} KL) and the latter encourages coverage of probability mass (known as \textit{forward} KL.) We exploit these two additional scores (denoted as \verb|KL-R| and \verb|KL-F|) with our Categorical CoDVAE model. Let be $x$ an input image and $\hat{x}$ its reconstruction. To ease the notation, we denote the set of pixel indices $P=\{i,j,k\mid i=0,\ldots,H-1,j=0,\ldots,W-1,k=0,\ldots,C-1\}$ for images of size $H\times W\times C$, and define our reconstruction-based scores on that as follows:
\begin{itemize}
    \item $\bce(x,\hat{x}) = -\sum_{p\in P} x_p\log \hat{x}_p+(1-x_p)\log 1-\hat{x}_p$. It denotes the sum of the \textit{binary-cross entropy} between the true and predicted pixels. Since the pixel values in each mask image are either zero or one, this measure is well-defined, without any normalization.
    \item $\sse(x, \hat{x}) = \sum_{p\in P}(x_p - \hat{x}_p)^2$. It represents the \textit{sum of squared errors} between the true and reconstructed pixels.
    \item $\sae(x, \hat{x}) = \sum_{p\in P}|x_p - \hat{x}_p|$. It depicts the \textit{sum of absolute errors}, which is the absolute difference between true and predicted pixels. It is worth noticing that if the Categorical CoDVAE model is evaluated on mask images, this metric is equal to the \verb|SSE| since both squared and absolute differences can be either zero or one (according to whether the pixel is correctly predicted or not): this is a direct consequence of taking the mode of the learned Bernoulli decoder, whose output pixels are binary values. Thus, in such cases, we omit the results of the SAE scores. Moreover, we can define $\sae$ on mask images too: $\sae$-mask$(x,\hat{x})=\sum_p\big|1[x_p>0]-1[\hat{x}_p>0]\big|$.
    \item $\dice(x,\hat{x}) = \frac{\sum_{p\in P} x_p^2 + \sum_{p\in P} \hat{x}_p^2}{2\sum_{p\in P}x_p\cdot \hat{x}_p}$. This score is defined as the inverse of the Dice coefficient \cite{deng2018learning}, to provide a measure of dissimilarity between two sets of pixels: $x$ and $\hat{x}$.  
    \item $\pixelsum(x,\hat{x}) = \sum_{p\in P}\hat{x}_p$. It is just defined as the sum of each predicted pixel value. This score can be interpreted as \textit{predicted total energy} if the model is trained to reconstruct $I$ (energy image), or as an approximation of the \verb|nTracks| if reconstructing $I_m$ (mask image), instead. We also investigated the total number of predicted non-zero pixels, i.e., $\sum_p 1[x_p > 0]$, but we did not report results about it because its discriminatory capability is quite weak.
\end{itemize}

\noindent Both categories of anomaly scores have their pros and cons. Reconstruction scores are in general easier to define, for example from either common loss functions or metrics, but are slower to compute them since it is required to forward the full model (encoder and decoder) to reconstruct the samples. Instead, latent-based scores involve only the encoder predictions which are more suited for inference, although possibly more difficult to define (e.g., analytical or empirical KL) and visualize (e.g., dimensionality reduction on high-dimensional latent space.)

\subsection{Training Procedure and Model Acceleration} 
\noindent During training, we make use of the set of augmentation functions, $\mathcal{T}$, defined in section \ref{subsec:data_aug}. At each mini-batch of mask images $I_m$ a random augmentation is sampled $t\sim \mathcal{T}$ and applied to it. The augmented images, $\Tilde{I} = t(I_m)$, are then fed to both residual and mask encoders; the decoder is then trained to reconstruct $\Tilde{I}$. The data augmentations enable the CoDAE models to learn invariances related to the coordinates $\eta$ and $\phi$ of the detector{, which can be also seen as an implicit way to impart some physics notions to the model}. In addition, we perform model selection according to the value of the \textit{structural similarity} (SSIM) \cite{DBLP:journals/tip/WangBSS04} metric between the true and reconstructed images, which provides a more human-aligned measure of image quality, computed on a validation set of only background samples.

The whole CoDAE models are learned end-to-end using the \texttt{AdamW} optimizer \cite{DBLP:journals/corr/KingmaB14}, whose learning rule decouples the weight decay regularization term from the main objective \cite{DBLP:conf/iclr/LoshchilovH19} making it easier to tune, minimizing the binary cross-entropy loss\footnote{The loss is summed over spatial dimensions (height, width, and channels), and averaged over the batch size.}. The optimizer is left with default parameters and learning rate, except for the weight decay coefficient set to $10^{-4}$. Furthermore, to improve training stability, we limit the $l_2$-norm of each gradient to be at most one. Lastly, all the weights are initialized by following the \texttt{he\_uniform} \cite{DBLP:conf/iccv/HeZRS15} scheme with zeros biases except for the decoder, whose biases are initialized to $-1$ to provide a better starting point for the initial reconstructions.

We use \textit{TensorFlow lite} \cite{DBLP:conf/osdi/AbadiBCCDDDGIIK16} for both model compression and acceleration of our CoDAE models. In particular, we employ a very light \verb|float16| weight and activation quantization that resulted in a $5.7\times$ reduction in model size and lower inference time, without any reduction in both anomaly detection and reconstruction performance. In this way we are able to achieve inference latency on a single consumer CPU hardware of about: $3$ms (i.e., $6-8\times$ faster) for the mask encoder, and $40$ms (i.e., $3.2 - 4.1\times$ faster) for a full forward pass on the whole model; therefore well under the $100$ms time-limit of the HLT. This also demonstrates that our model architecture is very easy to optimize for deployment.

\section{Results}
\noindent In this section we present our evaluation protocol, and show the obtained results of our experiments comparing physics-motivated baselines, prior approaches, and our models.

\subsection{Evaluation Metrics}
\label{subsec:metrics}
\noindent For the evaluation of both baselines (defined in the following section) and our proposed models, we treat the anomalies (the two signals) as the positive class and employ two popular metrics in out-of-distribution detection \cite{DBLP:conf/iclr/HendrycksMD19}: the area under the receiving operating characteristic curve (AUROC), and the false positive rate at $N\%$ of the true positive rate (FPRN). The AUROC summarizes the performance of the discriminator across multiple thresholds while the FPRN evaluates the performance at one specific threshold value: such threshold is often specific for the application and domain requirement. In our case, we choose $N\% = 40\%$ which targets a signal efficiency of $40\%$. Consequently, the metric is named FPR$40$. 

The AUROC can be considered {as} the probability that a signal sample is assigned a higher AD score than a background example. Thus, higher AUROC values are better, depicting a higher retention of the signal at a lower background efficiency (and so at a higher rejection rate of the background.) The FPRN metric, instead, is more suited to compare strong models: interpreted as the probability that a normal (background) sample is flagged as an anomaly (so as a signal) when the $40\%$ of signals are correctly detected. Since we want to decrease such false alarm probability, lower FPR$40$ values indicate a better model.

Moreover, since the performance of deep neural networks are dependent on many stochastic factors (e.g., sampling of the data, random weight initialization, dropout, etc) and specific (architecture and optimization) hyper-parameters, making difficult to conclude which model is actually better than another according to few (average) scores, especially when these show counter-intuitive results: e.g., one score is the best on a signal, and the worst on the other. Hence, we employ the \textit{Almost Stochastic Order} (ASO) test \cite{del2018optimal,dror2019deep} as implemented by \cite{ulmer2022deep} which provides a statistically significant result from which it can be decided the best performing algorithm. The ASO test is specifically designed for neural networks, building on the concept of \textit{stochastic order} in which one distribution of scores (e.g., AUROC) is said to stochastically dominate another one if the cumulative distribution of the former is lower than the latter for every point. Since the stochastic dominance is too strict to be practical, the \textit{almost stochastic dominance} is used instead, which quantifies the extent to which stochastic order is violated. Therefore, the ASO test returns an upper bound, called $\epsilon_{\min{}}$, expressing the amount of violation: if $\epsilon$$_{\min{}}$$<\tau$ (where $\tau$ is the \textit{rejection threshold} usually set to $0.5$ or less, like $0.2$ for a more confident result), then the former algorithm is superior than the latter. The value $\epsilon_{\min}$ can be interpreted as a confidence score: the lower it is, the more sure we can be about the dominance of one model over another. Experimentally, we follow the best practice suggested in \cite{ulmer2022deep} by fixing one set of hyperparameters and comparing multiple runs of the same model where possible. Moreover, for each benchmark mass we build an empirical distribution of the AUROC by computing this metric on a thousand of random, class-balanced subsets (with $2k$ samples each) of the data.

\subsection{Baseline Discriminators}
\label{subsec:baselines}
\noindent For a fair comparison and assessment of our method, we determined {various} baseline discriminators: two physics-motivated ones, a fully supervised classifier, a convolutional auto-encoder (CAE), and two anomaly detection models. The two physics discriminators are respectively based on the \textit{total energy} (i.e., the sum of energy deposits, $E_T$, in each image channel), and the \verb|nTracks| variable. Specifically, the \verb|nTracks| is a model-independent classical variable corresponding to the total number of tracks per event \cite{collaboration2008atlas,CMS:2008xjf}. Such quantity is related to the number of decay products providing the best approximation of such to discern the signal particles, being also independent of the binning used to discretize the detector resolution.
In particular, we define the total energy baseline as follows:
\begin{equation}
    \label{eq:tot_energy}
    s^{(k)}(c) = \sum_{i}^H\sum_j^W x_{i,j,c}^{(k)},\quad c\in\{0, 1, 2\}
\end{equation}
where $s^{(k)}(c)$ is the score value for channel $c$ (which denotes, respectively, the Trk, ECAL, and HCAL) of the $k$-th image $x^{(k)}$ with height $H$ and width $W$. Discrimination will be then performed according to the obtained scores, $s$, per channel.

Furthermore, we also consider the performance of a \textit{supervised} classifier, therefore assuming an ideal setting in which we would have perfect knowledge of the data. Such a supervised baseline will provide a good approximation regarding upper-bound discrimination performance that our unsupervised model may achieve at its best. In particular, we consider a robust model, a \textit{Compact Convolutional Transformer} (CCT) \cite{DBLP:journals/corr/abs-hassani21} that already outperformed a simpler convolutional network in our prior experiments.

The next baseline is a convolutional auto-encoder derived from our CoDAE: it has the same architecture and hyperparameters, lacking only the second (smaller) encoder network and the spatial conditioning mechanism in the decoder since there is no additional input (i.e., CoDAE's the auxiliary latent space) to perform conditioning on. This CAE model is trained in exactly the same way our models are.

The last baselines are two popular anomaly detection models: an unsupervised AE inspired\footnote{To the best of our knowledge, the authors provide only a figure outlining their model architecture and not a comprehensive description; we did our best to mimic their approach.}{by \cite{Heimel:2018mkt}, and the Dirichlet VAE from \cite{Dillon:2021nxw}. In particular, the AE model has a total of $600$k parameters, a latent space of size $32$, and was trained to minimize a mean squared loss: compared to our CoDAE, it lacks the second encoder and skip connections, and embeds its inputs to dense vectors while the images are reconstructed by transposed convolutions. Instead, the Dirichlet VAE is a weakly-supervised approach that, therefore, also requires a fraction of the signals for training: we assume a realistic setting in which the background is contaminated with $0.01\%$ of the signals. This model has a three dimensional latent space, whose prior distribution is the Dirichlet \cite{joo2020dirichlet} instead of the Gaussian, resulting in $2$M parameters: during training we normalize the images to sum to one, and use the same hyperparameters as in \cite{Dillon:2021nxw}. For both AD models, we apply the data augmentations defined in section \ref{subsec:data_aug}. We trained both AE and Dirichlet VAE for $50$ epochs (the latter converged earlier in training), the CoDAE for $30$ epochs, and the Categorical CoDVAE for $100$ of them since we observed slower convergence compared to the CoDAE. Lastly, the batch size is $128$ for all models, and the weights are optimized by} \verb|AdamW| \cite{DBLP:journals/corr/KingmaB14,DBLP:conf/iclr/LoshchilovH19}.

\subsection{Anomaly Detection}
{In this section we} compared our two models (the CoDAE and Categorical CoDVAE) against {the baselines defined in the previous section.} For all the models {and baselines} we computed the anomaly scores defined in section \ref{subsec:anomaly_scores}, where possible (e.g., it is not possible to compute the KL for the AE), and evaluated their respective AUROC and FPR40. In particular, in tables \ref{tab:results_auc} and \ref{tab:results_fpr} we provide the results for the best scores on average, {i.e., the anomaly scores that achieve the best performance on both signals by considering the average over the mass points}, while the full evaluation is {available} in the supplementary material.

\begingroup
    \setlength{\tabcolsep}{4pt}  
    \renewcommand{\arraystretch}{1.15}
    
    \begin{table*}[h]
        \centering
        
        \begin{tabular}{c|ccccccc|cccc}
            \toprule
            \multicolumn{1}{c}{\textbf{Model}} & \multicolumn{7}{c}{\textbf{SUEP (GeV)}} & \multicolumn{4}{c}{\textbf{SVJ (TeV)}} \\
              & 125 & 200 & 300 & 400 & 700 & 1000 & mAUC & 2.1 & 3.1 & 4.1 & mAUC      \\
            \midrule
            Total energy (\verb|Trk|) & 45.18 & 48.4 & 54.13 & 58.51 & 67.39 & 70.89 & 57.42 & 55.86 & 72.13 & 81.46 & 69.82 \\
            \verb|nTracks| & 78.68 & 92.39 & 98.31 & 99.6 & 99.94 & 99.93 & 94.81 & 82.05 & 89.32 & 92.92 & 88.1 \\
            Supervised CCT \cite{DBLP:journals/corr/abs-hassani21} & 89.72 & 96.58 & 98.88 & 99.42 & 99.87 & 99.93 & \textbf{97.4} & 96.05 & 98.17 & 98.76 & \textbf{97.66} \\
            \midrule
             *CoDAE ($Z_2$) & {79.77} & {93}  & {98.28} & {99.45} & {99.54}  & {99.23} & {94.88} & {83.2} & {88.3} & {90.83} & {87.44} \\ 
             *CoDAE (\verb|BCE|) & {88.74} &{97.54}  & {99.61} & {99.93} & {99.99} & {99.99} & \textbf{97.63} & {85.98} & {90.39} & {92.62} & \textbf{89.66} \\
             \midrule
            Cat. CoDVAE ($Z_1$) & 77.54 & 91.5 & 97.71 & 99.21 & 99.42 & 99.19 & 94.1 & 81.36 & 86.4 & 88.9 & 85.55 \\
            Cat. CoDVAE (\verb|KL-F|) & 69.32 & 83.92 & 93.18 & 96.38 & 98.04 & 98.18 & 89.84 & 79.69 & 84.08 & 86.11 & 83.3 \\
            Cat. CoDVAE (\verb|SSE|) & 86.93 & 97.01 & 99.51 & 99.9 & 99.98 & 99.98 & {97.22} & 85.29 & 89.78 & 92.04 & {89.04} \\
            \midrule
            *CAE (\texttt{BCE}) & {89.42}& {97.84}& {99.7 } &{99.95}& {99.99} &{99.99}& \textbf{97.81} & {85.96} &{90.45}& {92.72} & \textbf{89.71} \\
            AE \cite{Heimel:2018mkt}-like (\verb|PixelSum|) & 83.89 & 94.6 & 98.69 & 99.68 & 99.95 & 99.94 & 96.13 & 83.26 & 88.72 & 91.44 & 87.81 \\
            Dirichlet VAE \cite{Dillon:2021nxw} ($Z_2$) & 51.93 & 54.99 & 59.58 & 63.17 & 71.16 & 74.51 & 62.55 & 63.52 & 67.37 & 69.26 & 66.72 \\
            \bottomrule
        \end{tabular}
        \vspace*{0.2cm}
        \caption{Comparison of anomaly detection baselines and models on test-set, best scores only. AUROC metric, higher is better: mAUC denotes the AUROC averaged over all the mediator masses for a given signal. Entries associated to an (\texttt{*}) were averaged over three random seeds: $42$, $51$, and $73$. Top three best results in boldface.}
        \label{tab:results_auc}
    \end{table*}
\endgroup

\begingroup
    \setlength{\tabcolsep}{4pt}  
    \renewcommand{\arraystretch}{1.15}
    
    \begin{table*}[h]
        \centering
        
        \begin{tabular}{c|ccccccc|cccc}
            \toprule
            \multicolumn{1}{c}{\textbf{Model}} & \multicolumn{7}{c}{\textbf{SUEP (GeV)}} & \multicolumn{4}{c}{\textbf{SVJ (TeV)}} \\
              & 125 & 200 & 300 & 400 & 700 & 1000 & mFPR & 2.1 & 3.1 & 4.1 & mFPR      \\
            \midrule
            Total energy (\verb|Trk|) & 50.63 & 48.11 & 41.81 & 37.71 & 29.3 & 25.11 & 38.78 & 33.84 & 17.46 & 9.155 & 20.15 \\
            \verb|nTracks| & 11.84 & 2.849 & 0.277 & 0.028 & $\sim 0$ & $\sim 0$ & 2.5 & 5.45 & 1.599 & 0.539 & 2.53 \\
            Supervised CCT \cite{DBLP:journals/corr/abs-hassani21} & 0.242 & 0.016 & 0.002 & $\sim 0$ & $\sim 0$ & $\sim 0$ & \textbf{0.04} & 0.121 & 0.021 & 0.005 & \textbf{0.05} \\
            \midrule
             {*}CoDAE ($Z_2$) & {11.44} &  {2.68} & {0.35} & {0.107} & {0.237} & {0.434} & {2.54} & {4.65}& {2.006}& {1.17} & {2.61} \\
             {*}CoDAE (\verb|BCE|) & {5.832} & {0.806} & {0.057} & {0.004} &  {$\sim 0$} & {$\sim 0$}  & \textbf{1.12} & {2.829} & {1.027} &  {0.523} & \textbf{1.46} \\
             \midrule
            Cat. CoDVAE ($Z_1$) & 12.91 & 3.329 & 0.575 & 0.219 & 0.227 & 0.341 & 2.93 & 5.9 & 2.866 & 1.85 & 3.54 \\
            Cat. CoDVAE (\verb|KL-F|) & 19.08 & 7.498 & 2.635 & 1.452 & 0.739 & 0.677 & 5.35 & 7.425 & 4.913 & 4.085 & 5.47 \\
            Cat. CoDVAE (\verb|SSE|) & 6.652 & 0.944 & 0.071 & 0.007 & $\sim 0$ & $\sim 0$ & {1.28} & 3.065 & 1.176 & 0.588 & {1.61} \\
            \midrule
            {*CAE (\texttt{BCE})} & {5.175} & {0.654} & {0.04} & {0.002} &   {$\sim0$} &  {$\sim 0$}  & \textbf{0.98} & {2.694} & {0.943} & {0.464} & \textbf{1.37} \\
            AE \cite{Heimel:2018mkt}-like (\verb|PixelSum|) & 8.22 & 1.904 & 0.211 & 0.035 & $\sim 0$ & $\sim 0$ & 1.73 & 4.346 & 1.74 & 0.863 & 2.32 \\
            Dirichlet VAE \cite{Dillon:2021nxw} ($Z_2$) & 26.06 & 17.38 & 10.15 & 6.66 & 2.572 & 1.42 & 10.71 & 21.42 & 14.89 & 10.92 & 15.74 \\
            \bottomrule
        \end{tabular}
        \vspace*{0.2cm}
        \caption{Comparison of anomaly detection baselines and models on test-set, best scores only. FPR$40$ metric, lower is better: mFPR denotes the FPR$40$ averaged over all the mediator masses for a given signal. Entries associated with an (\texttt{*}) were averaged over three random seeds. Top three best results in boldface.}
        \label{tab:results_fpr}
    \end{table*}
\endgroup

Discussing about the physics baselines, for the SUEPs we can see that the sum of $E_T$ deposits of the signal is actually lower than the one of the QCD background, resulting in shifted distributions of scores that yield an AUROC below 50\%. This total energy baseline indicates that one signal (the SUEPs) is less complex than the background. Instead, such issue does not occur for the \texttt{nTracks} baseline, which already provides a pretty good separation performance for both signals on both evaluation metrics, especially for the SUEPs at high mass; as stated in \cite{Knapen:2016hky}: counting the number of tracks is particularly sensitive to high multiplicity soft particles like SUEPs, making them easier to identify. Compared to the AUROC, the FPR40 metric helps us better understand the shape of the ROC curve around our target signal efficiency of $40\%$, showing a very high (almost ideal) background rejection rate for the supervised classifier in all the benchmark scenarios.

\begin{figure}
    \centering
    \subfloat{\includegraphics[width=0.49\textwidth]{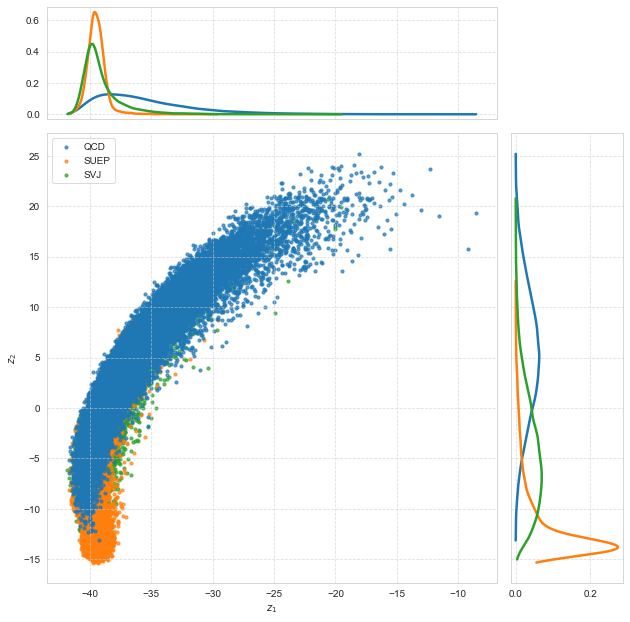}}
    \hfill
    \subfloat{\includegraphics[width=0.49\textwidth]{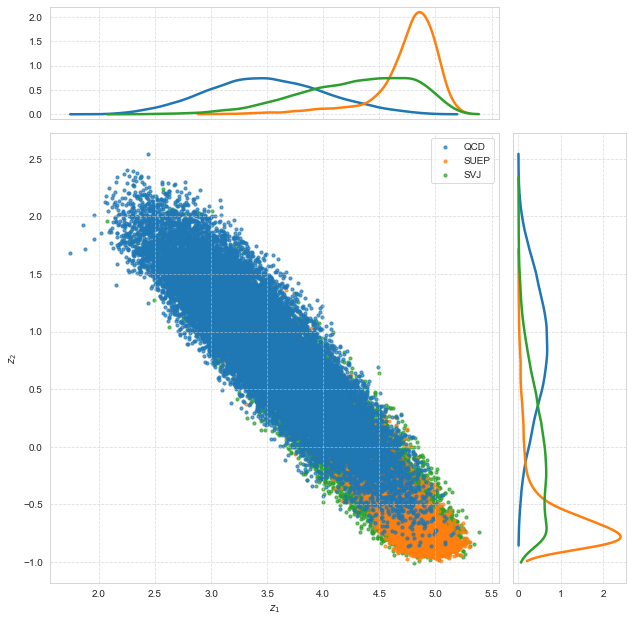}}
    
    \caption{The CoDAE's (left) and CoDVAE's (right) auxiliary latent space, $Z_m$, along with kernel density estimates of its components{: the best components are, respectively, $Z_2$ and $Z_1$}. As we can see {in the density plots}, {the} QCD (blue) and SUEPs (orange) look well separated, instead the SVJs (green) span in between the two classes.}
    \label{subfig:aux_z}
\end{figure}

Compared to the \verb|nTracks| baseline, which requires counting the track multiplicity possible only if fully reconstructing the event as in an off-line analysis, {our models} are already able to get competitive performance in the latent space ({showed in} figure \ref{subfig:aux_z}) and even improve by a neat margin when using reconstruction-based scores such as the \verb|BCE| or \verb|SSE|. We can notice that for some benchmark points, like \suep{400}, \suep{700} and \suep{1000}, the AUROC easily saturates (table \ref{tab:results_auc}) attaining an almost perfect background rejection (table \ref{tab:results_fpr}), therefore the improvement brought by a data-driven approach is negligible. Indeed, these are easier to detect since they are expected to deviate significantly from the QCD background. Instead, in the most challenging scenarios our best model achieves an AUC improvement of at most {$+10.1\%$} for SUEPs and {$+3.9\%$} for SVJs, as well as a reduction of FPR of at most {$6\%$} for SUEPs and {$2.6\%$} for SVJs, at the predefined signal efficiency compared to the \verb|nTracks| baseline. Our models are able to reduce the gap with the supervised classifier despite being trained on the background class only. {Next, the AE underperforms our models, and the Dirichlet VAE is only competitive against the total energy baseline attaining a too low background rejection rate. The CAE, then, performs slightly better than both the CoDAE and CoDVAE: this is expected since optimizing the model is easier, in fact, it can be seen as a simplified CoDAE as it lacks the spatial conditioning mechanisms.}

Lastly, in table \ref{tab:aso_test} we summarize the results about the ASO test between our models against the baselines, considering the AUROC metric as score distribution for the statistical test which is performed between two distributions at a time. As we can see, our models (in particular, when using the \texttt{BCE} and \texttt{SSE} scores) can beat most of the baselines and for the SUEP signal also the supervised classifier, except for the \suep{125} benchmark point. Furthermore, this test confirms the superiority of the CAE model over the CoDVAE but not against the CoDAE: the violation ratio fluctuates around $0.5$, making difficult to decide which model is the best. This also implies that the CAE is a valid alternative to our dual-encoder models, especially in settings in which the auxiliary latent space has a low discriminatory power.

\begingroup
    \setlength{\tabcolsep}{8pt}  
    \renewcommand{\arraystretch}{1.3}

    \begin{table*}[h]
        \centering
        \begin{tabular}{c|cccccc|ccc}
            \toprule
            \multicolumn{10}{|c|}{\textbf{*CoDAE ($Z_2$)}} \\
            \midrule
            \multicolumn{1}{c}{\textbf{Baseline}} & \multicolumn{6}{c}{\textbf{SUEP (GeV)}} & \multicolumn{3}{c}{\textbf{SVJ (TeV)}} \\
              & 125 & 200 & 300 & 400 & 700 & 1000 & 2.1 & 3.1 & 4.1      \\
            
            \midrule
            *CAE (\verb|BCE|) & 0.99 & 0.99 & 0.99 & 0.99 & 0.99 & 0.98 & 0.99 & 0.99 & 0.99 \\
            AE (\verb|PixelSum|) & 0.99 & 0.99 & 0.99 & 0.34 & 0.99 & 0.57  & 0.7 & 0.99 & 0.99 \\
            \verb|nTracks| & 0.66 & 0.65 & 0.99 & 0.99 & 0.99 & 0.99 & 0.33  & 0.99 & 0.99 \\
            Supervised CCT & 0.99 & 0.99 & 0.99 & 0.45 & 0.69 & 1  & 0.99 & 0.99 & 0.99 \\
            
            \toprule
            \multicolumn{10}{|c|}{\textbf{*CoDAE (\texttt{BCE})}} \\
            \midrule
            *CAE (\verb|BCE|) & 0.55 & 0.55 & 0.55 & 0.55 & 0.53 & 0.65  & 0.43 & 0.54 & 0.39 \\
            AE (\verb|PixelSum|) & 0 & 0 & 0 & 0 & 0 & 0  & 0 & 0 & $\sim0$ \\
            \verb|nTracks| & 0 & 0 & 0 & 0 & 0 & 0  & 0 & $\sim0$ & 0.99 \\
            Supervised CCT & 0 & 0 & 0 & 0 & 0 & 0  & 0.99 & 0.99 & 0.99 \\

            \toprule
            \multicolumn{10}{|c|}{\textbf{CoDVAE ($Z_1$)}} \\
            \midrule
            *CAE (\verb|BCE|) & 0.99 & 0.99 & 0.99 & 0.99 & 0.99 & 0.99  & 0.99 & 0.99 & 0.99 \\
        	AE (\verb|PixelSum|) & 0.99 & 0.99 & 0.99 & 0.99 & 0.99 & 0.99  & 0.99 & 0.99 & 0.99 \\
        	\verb|nTracks| & 0.99 & 0.99 & 0.99 & 0.99 & 0.99 & 0.99 & 0.99 & 0.99 & 0.99 \\
        	Supervised CCT & 0.99 & 0.99 & 0.99 & 0.99 & 0.99 & 0.99  & 0.99 & 0.99 & 0.99 \\

            \toprule
            \multicolumn{10}{|c|}{\textbf{CoDVAE (\texttt{KL-F})}} \\
            \midrule
            *CAE (\verb|BCE|) & 0.99 & 0.99 & 0.99 & 0.99 & 0.99 & 0.99  & 0.99 & 0.99 & 0.99 \\
        	AE (\verb|PixelSum|) & 0.99 & 0.99 & 0.99 & 0.99 & 0.99 & 0.99  & 0.99 & 0.99 & 0.99 \\
        	\verb|nTracks| & 0.99 & 0.99 & 0.99 & 0.99 & 0.99 & 0.99  & 0.99 & 0.99 & 0.99 \\
        	Supervised CCT & 0.99 & 0.99 & 0.99 & 0.99 & 0.99 & 0.99  & 0.99 & 0.99 & 0.99 \\

            \toprule
            \multicolumn{10}{|c|}{\textbf{CoDVAE (\texttt{SSE})}} \\
            \midrule
            *CAE (\verb|BCE|) & 0.99 & 0.99 & 0.99 & 1 & 0.99 & 0.99  & 1 & 1 & 1 \\
        	AE (\verb|PixelSum|) & 0 & 0 & 0 & 0 & 0 & 0  & 0 & 0 & 0 \\
        	\verb|nTracks| & 0 & 0 & 0 & 0 & 0 & 0  & 0 & 0 & 0.99 \\
        	Supervised CCT & 0.99 & 0 & 0 & 0 & 0 & 0  & 0.99 & 0.99 & 0.99 \\
            \bottomrule
            
        \end{tabular}
        \vspace*{0.2cm}
        \caption{Pairwise comparisons of models and baselines, performing the ASO test based on the AUROC with a $95\%$ confidence level: the headings are our models, which are compared to the baselines (the entries). Models and baselines denoted with a (\texttt{*}) were run on three random seeds. The smaller the value, the best the model compares against the respective baseline. Note: the listed values are the result of an approximate calculation, therefore there may be either false positives or negatives.}
        \label{tab:aso_test}
    \end{table*}
\endgroup

\subsection{Reconstruction Quality}
\noindent To assess the reconstruction quality of the compared models, we evaluate two metrics: the mean squared error (MSE), and the SSIM index \cite{DBLP:journals/tip/WangBSS04}. In particular, the popular MSE metric is useful for determining the texture quality of the reconstructions, since it penalizes pixel-level differences. Instead, the SSIM is a perceptual quality metric designed to better match the perceived visual quality of humans. These two metrics are complementary since the MSE looks at fine details while the SSIM at the global appearance of the images, providing a more comprehensive assessment of the quality of the reconstructed samples. 

\begin{figure*}[!t]
    \centering

    \begin{subfigure}{0.92\textwidth}
        \includegraphics[width=\textwidth]{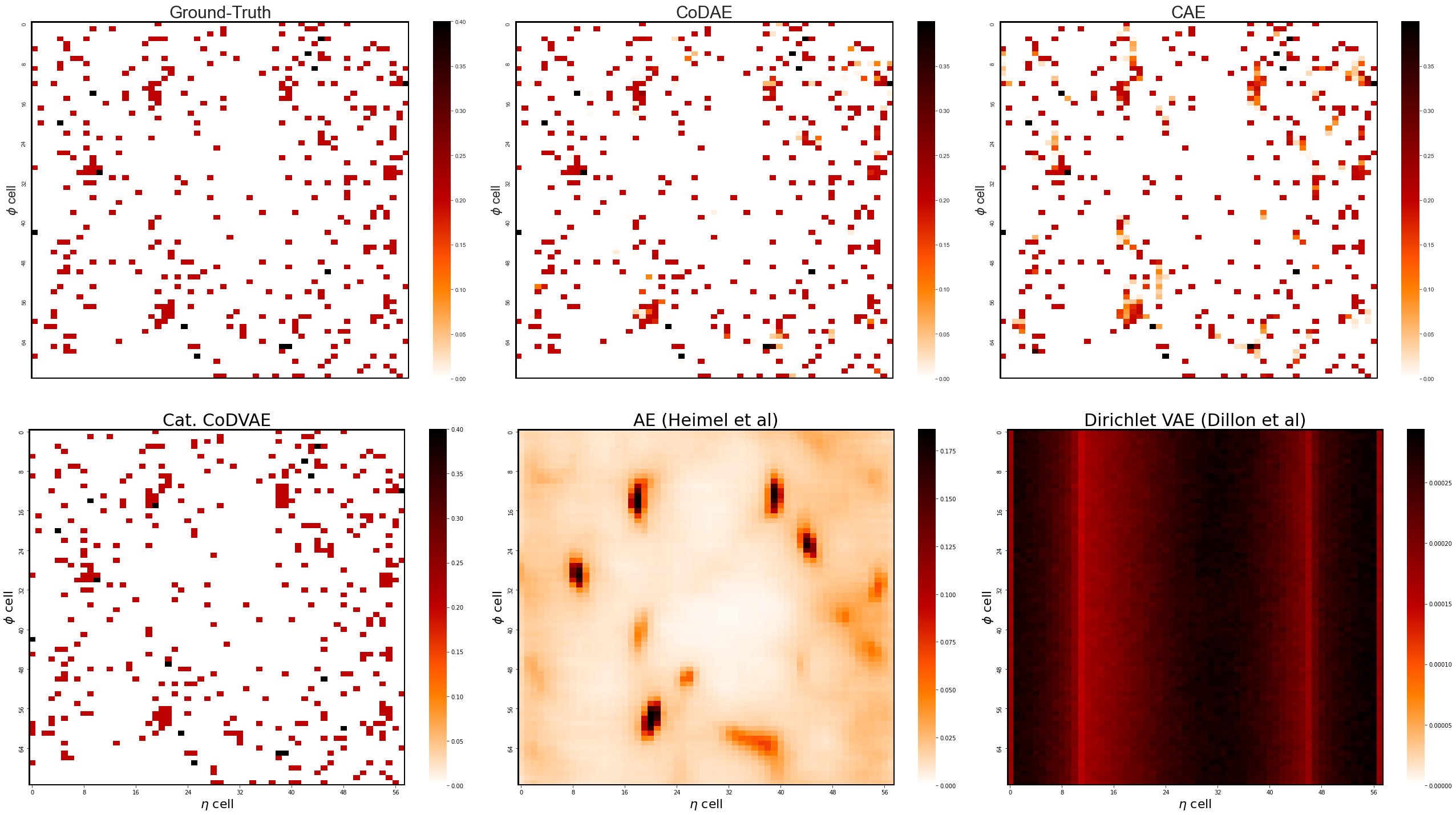}
        \caption{Reconstructions of five random QCD samples.}
        \label{subfig:per_sample_reco}
    \end{subfigure}

    \begin{subfigure}{0.92\textwidth}
        \includegraphics[width=\textwidth]{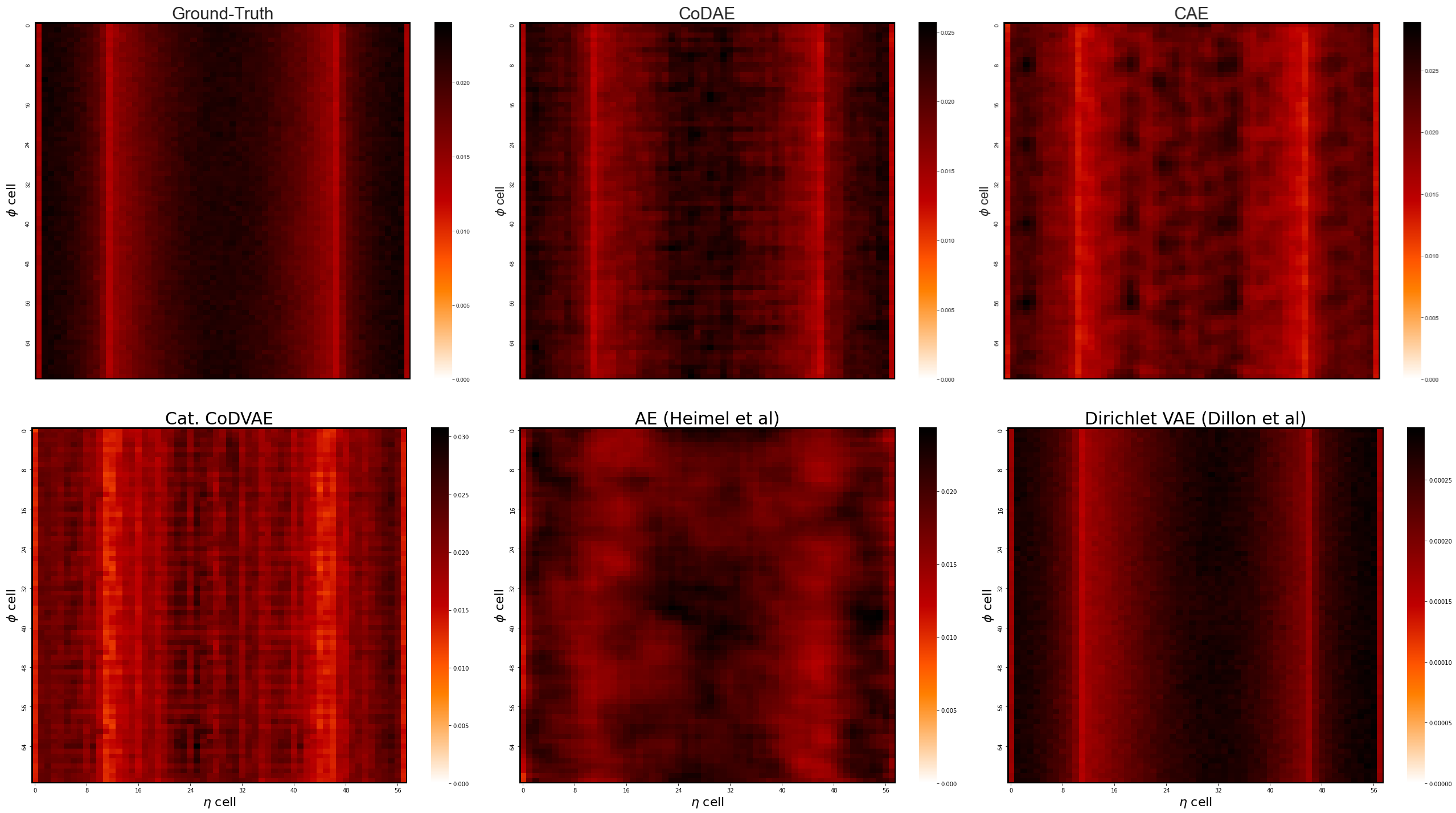}
        \caption{Reconstructions of the QCD test-set.}
        \label{subfig:avg_reco}
    \end{subfigure}
    
    \caption{Comparison of per-sample (a) and average ({b}) reconstructions. In order: ground-truth, CoDAE, {CAE,} Categorical CoDAE, AE \cite{Heimel:2018mkt}, and Dirichlet VAE \cite{Dillon:2021nxw}. As we can see our models are capable of pixel-accurate single-sample reconstructions, whereas the AE only captures the prominent pixels, and the Dirichlet VAE learns to predict a sort of average image instead. The CAE is also very accurate, as expected, since sharing the same encoder and decoder architecture of the CoDAE.}
    \label{fig:results_reco}
\end{figure*}

\begingroup
    \setlength{\tabcolsep}{2.6pt}  
    \renewcommand{\arraystretch}{1.25}

    \begin{table}[h]
        \centering
    
        \begin{tabular}{c|ccccc}
            \toprule
            \textbf{Metric} & \textbf{CoDAE} & {\textbf{CAE}} & \textbf{CoDVAE} & \textbf{AE \cite{Heimel:2018mkt}} & \textbf{Dirichlet VAE \cite{Dillon:2021nxw}} \\
            \midrule
            MSE & {\textbf{5.5 ($\pm$ 5.6)}} & {9.4 ($\pm$ 7.4)} & 17.7 ($\pm$ 13) & 70.9 ($\pm$ 24.7) & 79.3 ($\pm$ 26) \\
            SSIM & {\textbf{0.99 ($\pm$ 0.01)}} & {0.98 ($\pm$ 0.02)} & 0.95 ($\pm$ 0.04) & 0.36 ($\pm$ 0.12) & 0.17 ($\pm$ 0.11) \\
            \bottomrule
        \end{tabular}
        \vspace*{0.2cm}
        \caption{Evaluation of reconstruction quality: QCD test images. Lower values of MSE as well higher SSIM ones are better. Each entry denotes the average metric value as well as its standard deviation (in parenthesis.) By design the Dirichlet VAE outputs images that sum to one, so we undo the normalization before computing the metrics in order to have pixel values on the same scale. Best results are shown in boldface.}
        \label{tab:results_reco}
    \end{table}
\endgroup

\noindent Reconstruction performance is summarized in table \ref{tab:results_reco} as well visually in figure \ref{fig:results_reco}. From both we can deduce that our two models achieve the lowest MSE and highest structural similarity, attaining accurate single-sample reconstructions resulting in good predictions on average. As we can notice from table \ref{tab:results_reco}, all models show some variance in the reconstructed background, indicating that some QCD samples deviate from the ideal background event. Moreover, AE-based models show softer and smoother pixel predictions whereas the Categorical CoDVAE, thanks to its Bernoulli decoder, is capable of sharp reconstructions by taking the mode of each learned distribution, one per output pixel. Furthermore, in figure \ref{fig:results_reco} we can observe how the Dirichlet VAE, which is designed to capture a multi-modal latent space distribution, only captures the ``QCD mode'' since each sample, regardless of being background or signal, is predicted as a sort of average of QCD images: this is confirmed by the low reconstruction metrics. We noticed a similar behavior in our prior experiments when training regular AEs with a small latent space (e.g., $2$), even with a high-capacity residual encoder.

\noindent We want to highlight the importance of accurate (or at least coherent) reconstructions. Since anomaly detection scores can be defined from such predictions, it is necessary to avoid the model learning to predict some spurious pattern or artifact instead of the inputs, otherwise, it would be difficult to understand for a human expert why a new sample deviates from the training data. {Moreover, since an auto-encoder is trained to maximize the reconstruction quality which, in turn, mostly affects the reconstruction-based anomaly scores, it is equally important to avoid premature convergence of the model since this can lead to sub-optimal anomaly detection performance: employing a large latent space paired with powerful encoder and decoder networks can mitigate such issue, as seen for our CoDAE and CoDVAE.} This is essential to obtain coherent anomaly detection predictions, {high discrimination performance,} to have a trustworthy model, and even to debug the model itself once deployed.

\subsection{Discussion}
\noindent Throughout our study we showed how auto-encoders can be employed as an effective means to implement anomaly detection in HEP analyses. In particular, our models achieve higher background rejection rates meaning that the classified signal is less contaminated with false positives, which, in turn, boosts the statistical significance of actually finding a newly theorized physics signal when analyzing the real data collected during the LHC's runs{: for the SUEPs, we can even equal the supervised classifier even if this model still achieves largely superior rejection rates at the target signal efficiency.} In addition, our dual encoder architecture, which also learns a smaller network $f_m$, is particularly suitable for fast AD: the $f_m$ model can predict in just $3$ms, enabling more than real-time applications at the HLT, however at the cost of sacrificing accuracy{; in principle, such a model can be further optimized to match even tighter latency requirements, for example running on a FPGA hardware like in \cite{Govorkova:2021utb,valente2023joint}}. Moreover, our models just learn from raw detector images of particle collisions, requiring less or no effort to compute high-level variables (like counting the number of tracks) and objects (e.g., by a particle-based pre-processing), potentially simplifying the whole analysis setup. Lastly, our CAE model can be employed in scenarios in which the auxiliary latent space is not helpful and the latency requirements allow for a forward pass of the full model.

\section{Conclusions}
\label{subsec:ablations}
\noindent We demonstrate the first successful application of (variational) auto-encoders to deploy in the real-time event-triggering stages of experiments like ATLAS \cite{collaboration2008atlas} and CMS \cite{CMS:2008xjf} to search for two dark showers models: SUEPs and SVJs. Their discovery can potentially shed new light on the existence of dark matter and novel hidden sectors, which are currently uncovered and undercover at the LHC. 

Unlike the common trend in many related works \cite{Heimel:2018mkt, Cheng:2020dal,Dillon:2021nxw, Finke:2021sdf, dillon2023normalized}, we do not employ a specific particle-based pre-processing of our data, nor low- or high-level features, but instead learn directly from raw images of particle signatures obtained by discretizing the detector response, thus reducing the dependency on the physics model by only assuming tracking information to be relevant for the considered signals: although we discard the calorimeter information, we believe the ECAL and HCAL to be still useful in general, e.g., for searching long-lived particles.

Our models are evaluated against both signals, demonstrating anomaly scores that can identify both. Our CoDAE models aim to adapt to the background samples, potentially allowing us to generalize on whatever novel signal that is diverse from what the model learned about the QCD background. Ideally, it would be possible to train a single model to reject one or more background processes, filtering only the events that resemble a potential new physics signal. Our approach can potentially enable generic physics searches for unknown, new signals from raw images only and with little-to-no assumptions about the physics model. Further research and benchmark datasets would be required to fully accomplish such an important goal.

\clearpage

\bibliographystyle{ieeetr}
\bibliography{main}

\end{document}